\documentclass[a4paper,11pt]{article}
\usepackage{jheppub} 

\usepackage{amsmath}
\usepackage[dvipsnames]{xcolor}
\usepackage{mathrsfs}  
\usepackage{tikz}
\usetikzlibrary{decorations.pathmorphing}
\usetikzlibrary{decorations.markings}
\usepackage{pifont}
\usepackage{dsfont}
\usepackage{orcidlink}
\usepackage{comment}

\newcommand{\inu}[1]{\vec n_{\uparrow}^{(#1)} }
\newcommand{\ind}[1]{\vec n_{\downarrow}^{(#1)} }
\newcommand{\cA}{\ensuremath{\mathcal A} }
\newcommand{\cL}{\ensuremath{\mathcal L} }

\newcommand{\cO}{\ensuremath{\mathcal O} }

\preprint{IPPP/25/86, TIFR/TH/25-24}

\title{\boldmath The Potential of HEFT and the scale of New Physics}

\author[a]{Rodrigo Alonso \orcidlink{0000-0002-4357-8991}}
\author[b]{Susobhan Chattopadhyay \orcidlink{0009-0000-2346-2273}}
\author[a]{James Ingoldby \orcidlink{0000-0002-4690-3163}}

\affiliation[a]{Institute for Particle Physics Phenomenology, Durham University,
South Road DH1 3LE, UK}
\affiliation[b]{Tata Institute of Fundamental Research, Homi Bhabha Road, Colaba, Mumbai 400005, India}

\emailAdd{rodrigo.alonso-de-pablo@durham.ac.uk}
\emailAdd{susobhan.chattopadhyay@tifr.res.in}
\emailAdd{james.a.ingoldby@durham.ac.uk}

\abstract{We employ a geometric framework to compute the leading high-energy behaviour of tree-level scattering amplitudes in theories containing $N$ Nambu–Goldstone bosons and a single Higgs-like scalar with an arbitrary potential $V$. Using these methods, we obtain closed-form expressions for the leading contribution to the full infinite set of tree amplitudes involving any number of Goldstone and Higgs-like external states. These results are then used to derive total cross sections, decay rates, and perturbative unitarity bounds. We then apply our general formalism to the Standard Model scalar sector using the equivalence theorem, working in the regime of small field-space curvature, and use it to characterise the relation between Higgs Effective Field Theory (HEFT) and the Standard Model Effective Field Theory (SMEFT). We analyse three representative classes of models. Two reproduce previously known behaviours, while the third, based on a dilaton effective theory, exhibits a noteworthy feature within the HEFT framework: a smooth decoupling limit that connects HEFT directly to the Standard Model without passing through a SMEFT regime, providing a possible backdoor to the Standard Model.
}

\begin{document}
\maketitle
\flushbottom

\section{Introduction}
\label{sec:intro}

The Large Hadron Collider (LHC) at CERN has directly probed the electroweak symmetry-breaking (EWSB) scale and established that the Higgs doublet of the Standard Model is the primary source of EWSB. Crucial questions nevertheless remain: (i) is this doublet the only source of electroweak symmetry breaking, and (ii) are there additional, as yet undiscovered, states whose masses arise from the same mechanism as that of known matter? These are fundamental questions about possible ultraviolet (UV) completions of the SM, yet they can be translated into a model-independent language using Effective Field Theory (EFT). A key issue then becomes whether structurally different EFTs are required to describe qualitatively different possible UV completions of the SM at low energy.

From a physical standpoint, it can be argued why more than one EFT structure is required. For example, if the UV completion contains additional degrees of freedom whose masses arise from the electroweak vacuum expectation value (VEV), $v$, then a low-energy EFT that omits these states must necessarily become singular as $v\rightarrow0$, reflecting the appearance of new light modes. This structure is qualitatively different from the simpler possibility in which the EFT remains regular in the symmetry-restoring limit. This simple argument provides a starting point for the formulation of the two possible EFTs.

In practice, the Higgs VEV is not an externally tunable parameter, unlike order parameters in condensed matter systems. While cosmology may have provided such a dial in the early universe (see e.g.~\cite{Banta:2022rwg,Alonso:2023jsi,Kanemura:2022txx}), collider experiments probe the scalar sector indirectly through scattering processes involving a small number of external particles. These measurements constrain only the leading terms in an expansion of the EFT about the physical vacuum and therefore probe the scalar field space only locally. This makes the distinction between singular and non-singular behaviour as $v\rightarrow0$ challenging to establish experimentally in full generality.

Within this context, the Standard Model Effective Field Theory (SMEFT), formulated in terms of an electroweak doublet $H$, provides a particularly clear example of a theory with a decoupling limit: higher-dimensional operators are suppressed by a mass scale that can be taken arbitrarily large, and the symmetry-restoring point $H^\dagger H=0$ is assumed to be non-singular. This SMEFT case is also referred to as the linear realisation, following the classification in~\cite{Callan:1969sn,Coleman:1969sm}. Scenarios in which additional electroweak-scale states do not decouple instead give rise to EFTs that lack such a regular fixed point. These theories cannot be described by an action analytic in $H$, but they can be captured within Higgs Effective Field Theory (HEFT), in which the Higgs degree of freedom is treated independently from the Nambu–Goldstone bosons associated with electroweak symmetry breaking. This case in contrast with SMEFT is referred to as the non-linear realisation.

A crucial structural observation is that SMEFT is not disjoint from HEFT. By an appropriate field redefinition, a complex scalar doublet may always be rewritten in terms of a singlet radial mode and a triplet of non-linearly transforming Goldstone fields\footnote{Another basis is the broken-gauge-invariant field one~\cite{Graf:2022rco,Alonso:2024usj} where fields only transform under the unbroken symmetry.}. As a result, every SMEFT can be expressed as a particular instance of HEFT, and one may formally write SMEFT$\subset$HEFT.  Whether a given HEFT is genuinely non-linear, or instead secretly equivalent to a SMEFT, cannot therefore be decided by a choice of field variables. Any meaningful criterion capable of distinguishing between these possibilities must be invariant under field redefinitions.

This requirement provides a strong motivation for the use of geometric methods in EFT. By formulating the scalar sector in terms of field-space geometry, one ensures invariance under arbitrary field redefinitions at every step. This perspective has recently led to a range of new insights, including connections to soft theorems~\cite{Cheung:2021yog,Derda:2024jvo,Cohen:2025dex}, renormalisation~\cite{Alonso:2017tdy,Alonso:2022ffe,Helset:2022pde,Assi:2023zid,Jenkins:2023rtg,Jenkins:2023bls,Li:2024ciy,Aigner:2025xyt,Assi:2025fsm}, fermionic interactions~\cite{Finn:2020nvn,Gattus:2023gep,Gattus:2024ird,Craig:2025uoc}, and positivity bounds~\cite{Remmen:2024hry,Chakraborty:2024ciu}. The connection between analyticity of the action and unitarity was first emphasised in this context in~\cite{Falkowski:2019tft,Chang:2019vez} and later reconciled with geometry in~\cite{Cohen:2020xca}. An early example of what we will later describe as \emph{backdoor} behaviour can already be found in~\cite{Alonso:2021rac}.

The present paper builds on these developments by using geometric methods to establish a direct connection between ultraviolet assumptions and infrared scattering observables. We first derive general results for high-energy amplitudes, cross sections, decay rates, and unitarity bounds, and then apply this framework to HEFT through three representative classes of models. Two reproduce behaviours that are already well understood. The third, based on a dilaton effective theory, is of particular interest: it provides an explicit example of a HEFT that admits a smooth decoupling limit to the Standard Model without first passing through a SMEFT regime, thereby illustrating that a HEFT description may be essential even for theories that depart only mildly from the SM.

A notational remark for the rest of this paper; for brevity when referring to HEFT in the following one is to understand HEFT-minus-SMEFT. This is indeed abuse of notation, given SMEFT$\subset$HEFT yet the context should make it clear. This  alternative use of terminology is widespread in the literature.

The paper is organised as follows. Section~\ref{sec:GeoandPot} introduces the scalar theory under consideration, its geometric formulation, and the derivation of leading high-energy amplitudes with arbitrary multiplicity. Section~\ref{sec:UnitRels} derives expressions for cross sections, decay rates, and unitarity constraints. Analytic results are presented in Section~\ref{sec:Analytic}. Applications to the three exemplar models are discussed in Section~\ref{sec:Models}. We conclude in Section~\ref{sec:Concl}. Technical details are collected in Appendix~\ref{sec:app}.

\section{Geometry and potential-generated amplitudes for any number of Higgses and Goldstones}
\label{sec:GeoandPot}

A local change of field-coordinates $\phi(\varphi)$ does have no effect in the $S$ matrix provided both fields excite the same quanta, i.e. $\langle 0|\varphi a^\dagger|0\rangle\neq0$ where $a^\dagger$ creates a particle excitation of $\phi$. The consequence of this observation on amplitudes, via  differential geometry, is that they are kinematic-dependent tensors $T$ projected by vierbeins $e$ into `flat' space, i.e. for a $n$-point amplitude of scalars with flavour marked by $a_i$ indexes
\begin{align}\label{eq:AisaTensor}
    \mathcal A_{a_1\dots a_n}=\left(\prod_{i=1}^n e_{a_i}^{\alpha_i}\right) T_{\alpha_1\dots \alpha_n}(\left\{s_{jk}\right\})\,,
\end{align}
with $s_{jk}=(p_j+p_k)^2$ and Greek indices transforming under coordinate changes $\phi(\varphi)$ as
\begin{align}
    T^\alpha&\to \frac{\partial \phi^\alpha}{\partial\varphi^\beta}T^\beta\,, &T_\alpha\to\frac{\partial\varphi^\beta}{\partial\phi^\alpha}T_\beta\,.
\end{align}

One has therefore that field derivatives are vectors under field transformations $d\phi^\alpha=\partial_\mu\phi^\alpha dx^\mu$ as it simply follows from the chain rule, but $\phi$ itself is not. The usual QFT expansion is based on $\phi$ and so does not display the tensor-vierbein form of amplitudes manifestly. This structure can be made explicit and computations significantly simplified if one uses Riemann normal coordinates. Let us apply this to a theory of $N+1$ scalars $\phi^\alpha$ with a potential $V$ and a kinetic term which contains the metric $g_{\alpha\beta}(\phi)$ while Riemann normal coordinates are denoted by $\varphi$ with the transformation $\phi(\varphi)$ given in e.g.~\cite{Hatzinikitas:2000xe} so that one has
\begin{align}
    \mathcal L_2=&\frac{1}{2} \partial_\mu \phi^\alpha g_{\alpha\beta}(\phi) \partial^\mu \phi^\beta-V(\phi)\,,\\
    =&\frac12\partial_\mu \varphi^\alpha\partial^\mu \varphi^\beta\left(g(0)_{\alpha\beta}+\sum_{n=2}\frac{1}{n!}\frac{n-1}{n+1}  [\nabla^{n-2}]_{i_3\dots i_n}R_{\alpha i_1i_2\beta }(0)\varphi^{i_1}\dots \varphi^{i_n}\right)\nonumber\\&-\sum_n\frac1{n!}\varphi^{i_1}\dots \varphi^{i_n}[\nabla^n V]_{i_1\dots i_n}+\mathcal O(R^2)\,,
\end{align}
where it is relevant to note that this theory has two sources for interactions, the potential and the curvature with relative strength $\nabla^{4+n} V$ and $E^{2}\nabla^n R$. The virtue of these coordinates is a covariant expansion which yields tensors for interaction vertexes and a vanishing connection around the vacuum. A prerequisite for computing an $n$-point amplitude in this framework is therefore the knowledge of the tensors above to order $n$, the next step being obtaining the Feynman rules $\langle 0|\mathcal L|\phi^n\rangle$, an exercise in symmetrisation and combinatorics. 

Finding the Feynman-vertex-tensors to arbitrary number of particles implies as  first step computing arbitrarily many covariant derivatives; while straightforward order by order this might not always be as tractable and provide as explicit formulas as ordinary derivatives do. Let us at this point narrow down the type of theory by assuming there exists a $SO(N+1)$ symmetry which acts on $N$ of the fields which are Nambu-Goldstone bosons (NGB) living in $S^N$ and denoted $\pi$, while the other scalar does not transform under $SO(N+1)$ and will be referred to as the singlet. It follows that symmetry restricts the form of the Lagrangian to
\begin{align}\label{eq:underlinehl}
    \cL =& \frac{1}{2}K^2(\underline h)(\partial \underline h)^2+\frac{1}{2}(\underline F(\underline h))^2(\partial \pi)^j\hat g_{ij}(\pi)(\partial\pi^i)-\underline V(\underline h)\,,\\
    \cL =& \frac{1}{2}(\partial h)^2+\frac{1}{2}(F(h))^2(\partial \pi)^j\hat g_{ij}(\pi)(\partial\pi^i)-V(h)\,,
    \label{eq:lheft}
\end{align}
where the inner metric $\hat g$ is that of the $S^N$-sphere (one need not specify its form) and the second line follows from the first after the change of variable
\begin{align}
    \frac{d\underline h}{d h}&=\frac{1}{K(\underline h)}\,, &  h&=\int_0^{\underline h} K(\underline h') d\underline h'\,,\label{eq:hbarhchange}
\end{align}
where the vacuum is at $h=\underline h=0$.
In the following we will stay in the $h$ frame unless otherwise specified and will use Greek letters to comprise all indexes and Latin letters starting at $i$ for NGBs. One has that the manifold is locally\footnote{Globally on the other hand the scalar manifold can be a priori any extension of its local shape, some UV-motivated examples are $\mathbb{R}^{N+1}$ and $S^{N+1}$.} $S^N\times \mathbb{R}$ and the question of the fixed point is the existence of some value of $h$ for which $S^N$ collapses to a point (locally the origin of polar coordinates). The connection and vierbein read
\begin{align}
    \Gamma_{jh}^i=&\delta^i_j\frac{F'}{F}\,, &\Gamma^{h}_{ij}&=-FF'\hat g_{ij}\,, & e^\alpha_a&=\begin{pmatrix}
        1&0\\0& F^{-1}\hat e^i_{i'}
    \end{pmatrix}\,,
\end{align}
where $\hat e^{i}_{a_i}\hat e^{j}_{a_j}\hat g_{ij}=\delta_{a_i a_j}$, the only other non-vanishing element not given by symmetrising the above is $\Gamma^{i}_{jk}$, which is obtained from $\hat g_{ij}$ solely and once more we do not need its explicit form.

As for the computation of covariant derivatives, let us here try and find a recursion relation that we can solve for.  The addition of a $h$ index on a tensor with $b$ NGB indexes reads
\begin{align}
T_{i_1i_2\dots i_bh} \equiv   \nabla_h T_{i_1i_2\dots i_b}=\partial_h T_{i_1i_2\dots i_b}-\sum_k\Gamma_{hi_k}^j T_{i_1\dots j\dots i_b}=F^b \frac{\partial}{\partial h}\frac{T_{i_1i_2\dots i_b}}{F^b}\,,
\end{align}
where we have simplified the expression given $\Gamma_{hi}^h=0$, $\Gamma_{hi}^j=-F'\delta^j_{\,i}/F$ and we use the same letter for the derivative of a tensor and the tensor itself with the understanding that the extra indexes on top of the original ones are derivatives. In this way one can append as many $h$ indexes to a tensor as desired which therefore provides the building blocks for arbitrary-Higgs emission amplitudes. A relevant instance is Higgs-derivatives of the Riemann tensor, first given in~\cite{Cohen:2021ucp},
\begin{alignat}{2}
& R_{ihjh}=-FF''\hat g_{ij}\,,
&\qquad
& R_{ijkl}= \left(\frac{1}{v^2}-(F')^2\right)F^2
(\hat g_{ik}\hat g_{jl}-\hat g_{jk}\hat g_{il})\,,\\
& R_{ihjh;h^a}= F^2\left(\frac{d}{dh}\right)^a\frac{R_{ihjh}}{F^2}\,,
&\qquad
& R_{ijkl;h^a}= F^4\left(\frac{d}{dh}\right)^a\frac{R_{ijkl}}{F^4} \,.
\end{alignat}
Appending an NGB index cannot quite be given in such a general way. To make headway let us assume the tensor depends only\footnote{This is the case for the present theory but also holds to leading order when explicitly $SO(N+1)$ breaking is small, as in the SM} on $\pi$ though $\hat g(\pi)$ , then one has
\begin{align}
    \nabla_jT_{i_1i_2\dots i_b}=\partial_jT_{i_1i_2\dots i_b}-\sum_k\Gamma_{ji_k}^l T_{i_1\dots l\dots i_b}-\sum_k\Gamma_{ji_k}^h T_{i_1\dots h\dots i_b}=FF'\sum_{k=1}^b\hat g_{ji_k} T_{i_1\dots h\dots i_b}\,,\label{eq:GBDer}
\end{align}
where the NGB-only covariant derivative $\hat \nabla=\partial_j+\hat \Gamma$ vanishes since $\hat\nabla\hat g=0$ and the sum is to be understood as relocating the $i_k$ NGB index to $\hat g_{ji_k}$ while replacing it with a $h$ label. 

Let us take each tensor in the sum on the right hand side of Eq.~\eqref{eq:GBDer} with a $h$ entry in place of a $NGB$ entry to be obtained from a Higgs-derivative of a $b-1$ NGB tensor. The alternative is that such tensor is the original undifferentiated one, but this would be the initial condition and starting point of our recursion relation so there is no loss of generality in the assumption. To be precise the last term in the sum of Eq.~\eqref{eq:GBDer} is given by $\nabla_h T_{i_1...i_{b-1}}=T_{i_1...i_{b-1}h}$, the next to last is given by
\begin{align}
    T_{i_1...i_{l}hi_{b}}=&T_{i_1...i_li_{b}h}+T_{i_1...i_lhi_{b}}-T_{i_1...i_li_{b}h}=T_{i_1...i_li_{b}h}+(\nabla_{i_b}\nabla_h-\nabla_h\nabla_{i_b})T_{i_1...i_l}\,,\\
    =&T_{i_1...i_li_{b}h}+\sum_{k=1}^l R^{\beta}_{\,\,\,\alpha_k i_bh} T_{i_1...\beta...i_{l}}\,,
\end{align}
which is, to leading order in our curvature expansion, the same element. The same holds true for $h$ in different positions which can be given by the iterative procedure demonstrated in the formula above.
Note that the specific Riemann tensor element appearing is in our basis proportional to $F''$. One can in this small curvature limit obtain recursion relations for an original 0-tensor $V$ as
\begin{align}
    \nabla_h V_{i_1i_2\dots i_b}&=F^b \frac{\partial}{\partial h}\frac{V_{i_1i_2\dots i_b}}{F^b}\,,
    \\ \nabla_jV_{i_1i_2\dots i_b}&=FF'\sum_k \hat g_{ji_k} F^{b-1}\frac{\partial}{\partial h} \frac{V_{i_1\dots i_b}}{F^{b-1}}+\mathcal O(R)\,.
\end{align}
The solution to these recursion relations is, neglecting the $\mathcal O(R)$ term above,
\begin{align}
    V_{i_1,\dots i_b \,h\dots h}=&\hat g_{i_1\dots i_N}F^b \left(\frac{d}{dh}\right)^a \left(\frac{F'd}{Fdh}\right)^{b/2}V(h)\equiv \hat g_{i_1\dots i_b}  \mathcal D_{a,b}V(h)\label{eq:RecRelSol}\,,
\end{align}
where we have introduced the derivative $\mathcal D$, and $\hat g_{i_1\dots i_b}$ is the symmetrisation of $b/2$ metric tensors built recursively from the equation above
\begin{align}
    \hat g_{i_1..i_{b}}=&\sum_{k=1}^{b-2}\hat g_{i_1\dots i_b\dots i_{b-2}}\hat g_{i_{b-1}i_k}+\hat g_{i_1\dots i_{b-2}}\hat g_{i_{b-1},i_b}\,,
\end{align}
where $b$ is necessarily even and the sum is to be understood as taking the $k$'th index (including the one in the last $\hat g$) and replacing it with $i_b$. The $b$-index $\hat g$ has $(b-1)!!$ terms, the first two instances read
\begin{align}
    \hat g_{ij}&=\hat g_{ij}\,, & 
    \hat g_{ijkl}&=\hat g_{ij}\hat g_{kl}+\hat g_{ik}\hat g_{jl}+\hat g_{il}\hat g_{jk}\,.
\end{align}
Contracting all indexes with inner-metric vierbeins amounts to the substitution $\hat g_{ij}\to \delta_{ij}$ which is the tensor that does appear in amplitudes.

The explicit formula for the $n^\text{th}$ covariant derivative obtained gives the Feynman vertex for the interaction of $n$ particles in Riemann normal coordinates; more in detail one would have the $\varphi^n\nabla ^n V(0)/n!$ term in the Lagrangian contracted as $i\langle 0|\varphi^n\nabla ^n V/n!|\varphi^n\rangle= i\nabla ^n V$ so that symmetrisation cancels the factorial. This term also provides the leading potential-generated term in a high energy expansion; for the $n$-point amplitude one has the contribution of this contact $n$-point term, then lower point vertexes connected by propagators and hence necessarily suppressed by powers of $1/s$ at high energy. Let us note that, were one to do the same computation not in Riemann-normal coordinates, obtaining this leading term would require summing over the leading contributions of each diagram, and the number of diagrams scales exponentially with the number of external states.

One can therefore use the expression for covariant derivatives here obtained  to rewrite the known~\cite{Cohen:2021ucp} $2\pi$-$n$-Higgs amplitudes in the limit of high energy and small curvature $R$ as
\begin{multline}
    \cA(\pi_i\pi_jh^{n}) = \delta_{i_1i_2}\left[\frac{d^n}{dh^n}\frac{F'd}{Fdh}V\right](0)-\left(s_{12}-\frac{2m^2_h}{n+1}\right)\left[\frac{d^{n-2}}{dh^{n-2}}\frac{R_{ihjh}}{F^2}\right](0) \\ 
    + \cO(\bar{V}\bar{R},\,\bar{R}^2)+\text{factorisable pieces.}
\end{multline}
but also genuinely new results for any number of NGB and Higgses
\begin{align}
    \mathcal A(\pi_{i_1}\dots \pi_{i_b},h^a)=\delta_{i_1\dots i_b}\left[ \left(\frac{d}{dh}\right)^a \left(\frac{F'd}{Fdh}\right)^{b/2}V\right](0)+\mathcal{O}(R)+\text{factorisable pieces.}\label{eq:newAmp}
\end{align}
which would be a good approximation to the actual amplitude if $E^n\nabla^{4+n} V$ terms dominate over $E^{n+2}\nabla^n R$, in other words the small curvature limit.  The factorisable pieces can be obtained recursively with the bootstrap approach from lower point amplitudes and have a rational form with poles for resonance-propagation; for our purposes their relevant property is that in the high energy limit their $s$ dependence is subleading to the contact-point-like terms here considered. For further details we refer the reader to~\cite{Cohen:2021ucp}.

The limit $R\to0$ yields naively a linear-sigma model with arbitrary potential $V(h)$, with an $N+1$-plet of scalars triggering $SO(N+1)\to SO(N)$ breaking. Let us stress however that the potential here can be non-analytic and have indeed a singularity at or even closer than the would-be symmetry restoring point which is why we will avoid the name linear-sigma model since this is not the same `linear' of a linear realisation.

\section{Infinite amplitude sums for unitarity, cross sections and decay rates}
\label{sec:UnitRels}

Unitarity constraints are the foremost theory tool to determine the edge of validity of current theories and hence signal the appearance of new physics. A given unitarity constraint is obtained from an element of the matrix equation $S^\dagger S=\mathds{1}$ and it contains itself a sum over elements of $S$; this sum, given $S$ is infinite dimensional, is itself infinite and hence lies the possibility of testing the theory at the non-local level. In the present approach this non-local information would be obtained from the full sum of terms in the covariant Taylor expansion around the vacuum. For a bottom up EFT approach, on the other hand, each term is {\it a priori} a variable and one has infinitely many parameters. Whether pitting these infinite number of constraints against the infinitely-many parameters results in a predictive outcome we do not know, yet it is the purpose of this section to provide the infinitely many constraints in a first step in this direction.

Deriving the unitarity constraints implies performing sums over intermediate states and integrals over phase space;  these results are connected to physical observables like cross sections and decay rates. In this section we will also point out this link and provide expressions for cross sections in terms of sums over infinitely many channels, formulas which we believe to be shown here for the first time.

\subsection{Perturbative unitarity constraints}

Unitarity demands, with our convention being $S=1-i\mathcal A(2\pi)^4\delta^4(p)$, that amplitudes satisfy
\begin{align} &2 \textrm{Im}(\mathcal A_{\textrm{in},\textrm{out}})+\sum_X \mathcal A_{\textrm{in},X}\mathcal A^\dagger_{\textrm{out},X}=0\,,
\end{align}
where `in' and `out' stand for the initial and final states, $X$ stands for intermediate particle states and their invariant Lorentz phase space with appropriate symmetry factors, and we have factored out and cancelled a Dirac delta for overall momentum conservation $(2\pi)^4\delta^4(p_{\textrm{in}}-p_\textrm{out})$ (i.e. we have selected only the fully connected pieces). One symmetry factor would be $1/2$ in the case in which the intermediate state consists of two Higgses, 
\begin{align}
    \frac12\int \frac{d^3p_1}{(2\pi)^32E_1}\frac{d^3p_2}{(2\pi)^32E_2}(2\pi)^4 \delta^4(p_\textrm{in}-p_1-p_2)\,.
\end{align}
If one adds an NGB there would be another factor of $d^3p/2E(2\pi)^3$ but still a $1/2$ in front, add two NGBs and now there is an extra $1/2$ if they have the same `flavour', else $1$; and so on and so forth. Finding the symmetry factor is the straightforward if seemingly painstaking generalisation of this. Geometry offers to take the pain away from this procedure as follows. The unitarity constraint is a tensor equation and the symmetry factor appears in the quadratic term which is made of a tensor times its hermitian conjugate tensor with $P$ of their indexes summed over, it reads 
\begin{align}
    \mathcal A_{\textrm{in},X}\mathcal A^\dagger_{\textrm{out},X}
    =&\int\mathcal A_{\textrm{in}\, a_1\dots a_P}\mathcal A_{\textrm{out}}^{\quad a_1\dots a_P}d\Phi_P
    =\int\mathcal A_{\textrm{in}\, \alpha_1\dots \alpha_P}\mathcal A_{\textrm{out}}^{\quad \alpha_1\dots \alpha_P}d\Phi_P\,,
\end{align}
where we have made the intermediate indexes explicit by substituting $X$ for a phase-space integral and sum over tensor indexes. The use of this equation is that, if the intermediate sum is truly a sum over tensor indexes containing all the information on particle species, they should also have the relative factors that account for degeneracies and give the sought after symmetry factor. They do in fact and one simply has to obtain the overall factor which is found to be 
 the phase space measure for $P$ \textit{identical} particles. The unitarity constraint reads in explicit form,
\begin{align}
    2\textrm{Im}(\mathcal{A}_{\textrm{in},\textrm{out}})
    +\sum_P\frac{1}{P!}\int(2\pi)^4\delta^4(p_\textrm{in}-\sum p)\left[ \prod_{k}^P\frac{d^3p_k}{(2\pi)^32E_k} \right]\mathcal A_{\textrm{in}\, a_1\dots a_P}\mathcal A_{\textrm{out}}^{\quad a_1\dots a_P}=0\,,
\end{align}
where each $a_i$ runs over all species of scalars.

In what follows we evaluate the amplitudes at tree level and retain only the leading growth with $s$, as in Eq.~\eqref{eq:newAmp}. These approximate amplitudes determine the parametric scale of the perturbative unitarity bound; higher-order corrections do not remove the growth of amplitudes and hence do not parametrically weaken the bounds. For amplitudes involving electroweak gauge bosons, the leading energy-growing contributions arise from longitudinally polarised external states. By the equivalence theorem, these are given by the corresponding Goldstone boson amplitudes up to corrections suppressed at large $s$. Amplitudes involving transverse vector bosons grow more slowly with energy and therefore do not affect the leading perturbative unitarity bounds.

The unitarity constraints are most conveniently expressed in a partial-wave decomposition, in which each partial wave satisfies an independent condition. In the present case, however, the amplitudes are momentum independent, so only the first partial wave is non-zero and the unitarity relation reduces to
\begin{align}
    2\textrm{Im}(V_{\textrm{in},\textrm{out}})
    +\sum_{P}\frac{\textrm{Vol}_{P}}{P!}V_{\textrm{in}\, \alpha_1\dots\alpha_P}V_{\textrm{out}}^{*\,\,\,\, \alpha_1\dots\alpha_P}=0\,,
\end{align}
 where Vol$_P$ denotes the $P$-distinguishable particle phase-space volume\footnote{Note that accounting for dimensionality of spacetime $d\neq 4$ is straight-forward to do in our derivation, the d-dimensional phase-space volume in particular is to be found in~\cite{Gehrmann-DeRidder:2003pne}.}
\begin{align}
    \textrm{Vol}_P=\frac{s^{P-2}2\pi}{(4\pi)^{2(P-1)}}\frac{1}{(P-2)!(P-1)!}\,.
\end{align}

\begin{figure}
    \centering
    \includegraphics[width=0.45\linewidth]{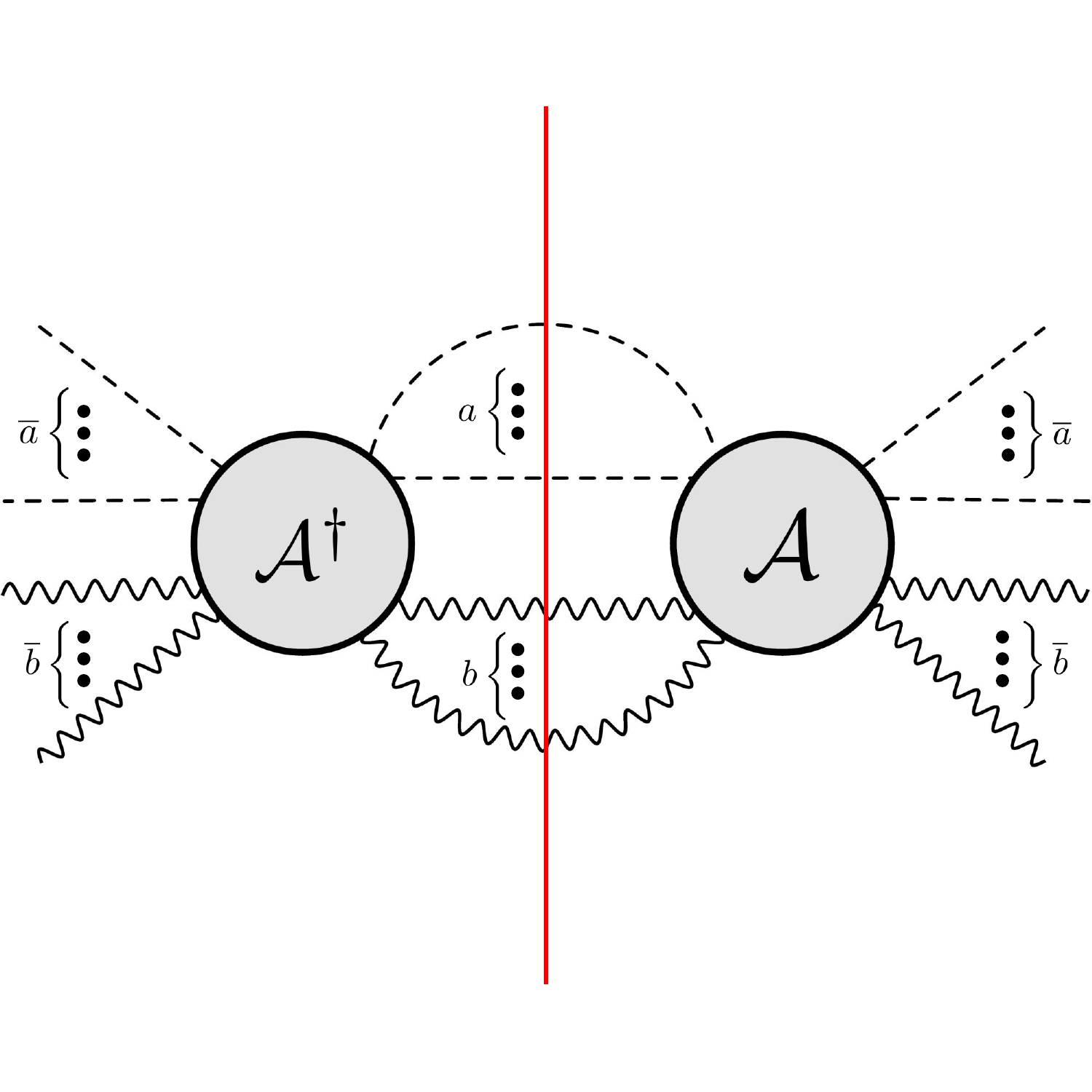}
    \caption{Diagrammatic depiction of the quadratic term in amplitude for unitarity constraints; we have $\bar a$ Higgs-like and $\bar b$ NGB's coming in and going out while the sum is over $a$ Higgs-like and $b$ NGB particles. }
    \label{fig:InOut}
\end{figure}
To further make these bounds explicit let us specify external and internal states, with a visual depiction provided in fig.~\ref{fig:InOut}. The in (out) state contains $\bar a$ ($\underline a$)-Higgses and $\bar b$ ($\underline b$) NGBs  whereas a given term in the tensor sum contains $a$ internal Higgses and $b$ NGB with $P=a+b$, $\bar P=\bar a+\bar b$. Given the symmetric properties of the tensors in our approximation in the internal sum, tensor elements with indexes that are a permutation of Higgs with NGB indexes  all add up as 
\begin{align}
    T_{\underbrace{ h\dots h}_{a\textrm{ times}} i_1..i_{b}}+T_{h i_1 \underbrace{h\dots h}_{a-1\textrm{ times}} i_2\dots i_{b}}+\dots= \frac{(a+b)!}{a!b!} T_{h\dots h i_1..i_{b}}\,,
\end{align}
and with our solution to the recursion relation in Eq.~\eqref{eq:RecRelSol} for small curvature the constraints read
\begin{align}
    &2\textrm{Im}(\mathcal D_{\bar a+\underline a,\bar b+\underline b}V)\delta_{i_1\dots i_{\bar b}i_{\bar b+1}\dots i_{\bar b+\underline b}}\nonumber\\+&\sum_{a,b}\frac{\textrm{Vol}_{a+b}}{a!b!}(\mathcal D_{\bar a+a,\bar b+b}V)(\mathcal D_{a+\underline a,b+\underline b}V)^*\sum_{\{j\}}\delta_{i_1\dots i_{\bar b}j_{1}\dots j_{b}}\delta_{i_{\bar b+1}\dots i_{\bar b+\underline b}j_{1}\dots j_{b}}=0\,.
\end{align}
Let us next take the same in and out state, which we do by setting $\bar a=\underline a$ and $\bar b=\underline b$ and also ensuring that the $k$ and the $k+\bar b$ NGB indexes are for the same flavour as one can do multiplying with Kronecker deltas, e.g. for the imaginary part
\begin{align}
    \left(\prod_{k=1}^{\bar b} \delta_{i_ki_{\bar b+k}}\right)\times 2\textrm{Im}(\mathcal D_{2\bar a,2\bar b}V)\delta_{i_1\dots i_{2\bar b}}\,.
\end{align}
Lastly we sum over external NGB indexes, using the result taken from~\cite{Alonso:2025vrn}
\begin{align}
    \sum \delta_{i_1 \dots i_b}\delta_{i_1 \dots i_b}=\hat g_{i_1 \dots i_b}\hat g^{i_1 \dots i_b}\equiv \mathscr{P}_b(N)=\frac{(b-1)!!(N+b-2)!!}{(N-2)!!}\,,
\end{align}
where we note that  $(-1)!!=1$. This procedure turns the tensor constraint into a scalar one as
\begin{align}
    2\textrm{Im}(\mathcal D_{2\bar a,2\bar b}V)\frac{\mathscr{P}_{2\bar b}}{(2\bar b-1)!!}+\sum_{a,b}\frac{\textrm{Vol}_{a+b}|\mathcal D_{\bar a+a,\bar b+b}V|^2 \mathscr{P}_{b+\bar b}}{a!b!}=0\,.
\end{align}
The choice of summing NGB indexes simplified the unitarity condition to the commonly encountered form but we do not purport to have taken the strongest of possible projections in tensor indexes, the study of these other possibilities is left for future work.

The constraint we have obtained takes the usual form, in the complex Argand plane defined as $\mathcal D_{2\bar a,2\bar b}V=x+iy$ , it can be cast in the form
\begin{align}
    r^2(1-\kappa)=x^2+(y+r)^2\,,
\end{align}
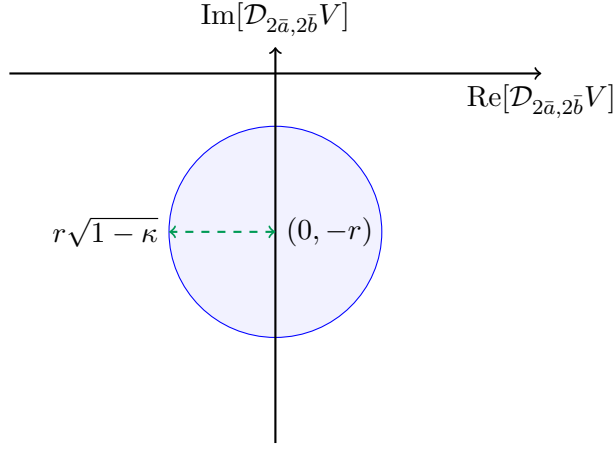
\begin{figure}
    \centering
    \begin{tikzpicture}[scale=0.7]
        \filldraw [blue,fill=blue!5] (0,-3) circle  (2);
        \draw [thick,->] (-5,0) -- (5,0) node [anchor=north] {Re$[\mathcal D_{2\bar a,2\bar b}V]$};
        \draw [thick,->] (0,-7) -- (0,0.5) node [anchor=south] {Im$[\mathcal D_{2\bar a,2\bar b}V]$};
        \draw [thick,dashed,ForestGreen,] (0,0) (0,-3) node [anchor=west,black] {$(0,-r)$};
        \draw [thick,dashed,ForestGreen,<->] (0,-3) -- (-2,-3) node [anchor=east,black] {$r\sqrt{1-\kappa}$};
    \end{tikzpicture}
    \caption{Unitarity constraint in the Argand plane; the real and imaginary parts of the $2\bar a$-Higgs and $2\bar b$ NGB derivatives of the potential should lie within the blue shaded circle of radius $r\sqrt{1-\kappa}$.\label{fig:Argand}}
\end{figure}
which confines $(x,y)$ to within a radius $r\sqrt{1-\kappa}$ of the point $(0,-r)$ as shown in fig.~\ref{fig:Argand} with
\begin{align}
    r\equiv \frac{\bar a!\bar b!}{(2\bar b-1)!!\textrm{Vol}_{\bar a+\bar b}}\,,
\end{align}
and 
\begin{align}
    \kappa_{\bar a,\bar b}&=\sum_{a,b\neq \bar a ,\bar b}\frac{((2\bar b-1)!!)^2\textrm{Vol}_{\bar a+\bar b}\textrm{Vol}_{a+b}\mathscr{P}_{\bar b+b}}{ a!b!\bar b!\bar a!\mathscr{P}_{2\bar b}}|\mathcal D_{a+\bar a,b+\bar b}V|^2\,,\\
    &=\sum_{a,b\neq \bar a ,\bar b}\frac{(2\bar b-1)!!(b+\bar b-1)!!(N+b+\bar b-2)!!(2\pi)^2s^{\bar P+P-4}|\mathcal D_{a+\bar a,b+\bar b}V|^2}{a!b!\bar b!\bar a!(N+2\bar b-2)!!(\bar P-1)!(\bar P-2)!(P-1)!( P-2)!(4\pi)^{2(P+\bar P-2)}}\,,
\end{align}
which is an explicit form of $\kappa$ that can readily be used to set bounds on the theory. 

Let us provide another form of $\kappa$ by introducing dimensionless magnitudes
\begin{align}
   w=&\frac{s}{(4\pi v)^2}\,, & \overline{\mathcal D}_{a,b}=&v^{a+b} \mathcal D_{a,b}\,, & \mathcal V&=V/v^4\,,
\end{align}
while defining
\begin{align}
    C_{\bar a,\bar b}&\equiv\frac{(2\bar b-1)!!(\bar a+\bar b-1)}{64\pi^2\bar a!\bar b!(\bar a+\bar b-1)!^2(N-2+2\bar b)!!}\,,
\end{align}
to recast the double sum into a nested sum over, first, all values for number of Higgs and NGB particles that add up to $P$, and then summed over $P$ itself, 
\begin{align}\label{eq:master}
     \kappa_{\bar a,\bar b}(w)&=C_{\bar a,\bar b}w^{(\bar a+\bar b-4)}\sum_{P}^\infty \frac{(P-1)w^{P}}{(P-1)!^2}\alpha_P[\mathcal V]-\left|\frac{\mathcal D_{2\bar a,2\bar b} V}{r}\right|^2\,,\\
     &=C_{\bar a,\bar b}\left[\sum_{P}^\infty \frac{(P-1)w^{P+\bar P-4}}{(P-1)!^2}\alpha_P[\mathcal V]-\frac{(\bar P-1)w^{2\bar P-4}}{(\bar P-1)!^2}\bar\alpha_{\bar a, \bar b}[\mathcal V]\right]\label{eq:master2}\,,
    \end{align}
where $\bar P=\bar a+\bar b$ and 
    \begin{align}
      \alpha_P[V]&=\sum_{B}\frac{(2B+\bar b-1)!!(2B+\bar b+N-2)!!}{(P-2B)!(2B)!} \left|\overline{\mathcal D}_{\bar a+P-2B,\bar b+2B}\mathcal V\right|^2\label{eq:alphaV}\,,\\
      \bar \alpha_{\bar a,\bar b}[V]&=\frac{(2\bar b-1)!!(2\bar b+N-2)!!}{\bar a!\bar b!}\left|\overline{\mathcal D}_{2\bar a,2\bar b}\mathcal V\right|^2\,,\\
         \frac{\mathcal D_{2\bar a,2\bar b} V}{r}&=\frac{w^{\bar a+\bar b-2}(2\bar b-1)!! \overline{\mathcal D}_{2\bar a ,2\bar b}\mathcal V}{8\pi \bar a!\bar b! (\bar a+\bar b-1)!(\bar a+\bar b-2)!}\,.
      \end{align}
with $b=2B$ and the sum over $B$ such that the total number of NGBs $\bar b+b$ is an even number. This leads to the distinction
\begin{align}
      d_{\bar b}&=2(\bar b/2-\lfloor \bar b/2\rfloor)=\left\{\begin{array}{cc}
           0& \textrm{if }\bar b \textrm{ even}  \\
           1 & \textrm{if }\bar b \textrm{ odd ,}
      \end{array}\right.\\ \sum_B=&\sum_{B=d_{\bar b}/2}^{\lfloor (P-d_{\bar b})/2\rfloor+d_{\bar b}/2}=\left\{\begin{array}{cc}
           \textrm{over positive integers+0}&\textrm{if }\bar b \textrm{ even}  \\
           \textrm{over positive semi-integers } & \textrm{if }\bar b \textrm{ odd .}
      \end{array}\right.
\end{align}

A number of remarks are in order concerning the $\kappa$ functions
\begin{itemize}
    \item The sum starts at $P=2$, the $P=1$ term cancels. This follows from our assumptions of high centre-of-mass energy squared $s$ with particle masses vanishing, which means one can never hit the pole that would make the one-particle exchange contribute.
    \item For scattering we further have the out states $\bar a+\bar b\geq 2$ which means the first power of $w$ in $\kappa_{\bar a,\bar b}$ is $w^{\bar a+\bar b-2}$, i.e. a constant for $2\to X$, $w$ for $3\to X$ etc.
    \item The $\kappa$ functions are given for an arbitrary number of external states labelled by $\bar a,\bar b$ and hence constitute an infinite set of bounds. For the cases studied in this work, a higher number of external states does not generically lead to stronger bounds as shown in a certain approximation in sec.~\ref{sec:AnalyGB}. It is sometimes the case however that a theory has $\kappa$ vanishing for low multiplicity external states, and for the first non-trivial bound one has to go to high multiplicity states.
    \item All terms in the sum are positive-definite with $n_{\bar P}\geq0$ powers of centre-of-mass energy; for theories valid to arbitrarily high energies one should have kappa functions that asymptote to some value smaller than one. The renormalisable case attains this by the `trivial' instance of having only $2\to 2$ contributions (and hence constant) while all others decrease with energy. 
    \item For finite energy each term within $\alpha_P[V]$ should be substituted as
    \begin{align}
        \left|\overline{\mathcal D}_{\bar a+P-2B,\bar b+2B}\mathcal V\right|^2\to\left|\overline{\mathcal D}_{\bar a+P-2B,\bar b+2B}\mathcal V\right|^2 \Theta\left(\sqrt{s}-P\,m_h-2B (m_\pi-m_h)\right)\,,\label{eq:Heavy}
    \end{align}
    where $m_h$ and $m_\pi$ are the masses of the Higgs and NGB respectively; mass terms have been omitted so far since they are taken to be small, i.e. $s\gg m_i$, yet this requires qualification. For the model discussed so far, c.f. Eq.~\eqref{eq:lheft}, the symmetry is global and exact so $m_\pi=0$ but in application to the e.g. HEFT one would equate $m_\pi$ with the $W$ and $Z$ masses. Even in the global symmetry limit, only channels with $a=P-2B$ Higgs-like particles such that $s> (am_h)^2$ are accessible and this is a finite set for finite $s$. The difference between a finite and infinite sum is assessed by the convergence of the series; for the problem at hand and as we will show, the convergence is good and the error in summing over all channels is small.
    
    \item The case $\bar a+\bar b=1$ can be connected to a physical decay rate if one substitutes $s=m^2$ (i.e. $w=m^2/(4\pi v)^2$) and restricts the sum in Eq.~\eqref{eq:alphaV} as would follow from including mass thresholds as in Eq.~\eqref{eq:Heavy}; we shall show this momentarily in Section~\ref{sec:XsecDecayR}.

\end{itemize} 

The set  of constraints obtained on the real part of the elastic amplitude read
\begin{align}
    \left[\frac{(2\bar b-1)!! \textrm{Re}(\overline{\mathcal D}_{2\bar a,2\bar b}\mathcal V)w^{\bar a+\bar b-2}}{8\pi\bar a!\bar b!(\bar a+\bar b-1)!(\bar a+\bar b-2)!}\right]^2&\leq  1-\kappa_{\bar a,\bar b}\left(\overline{\mathcal D}\mathcal V\,;w\right)\,, \label{eq:BonReal}
\end{align}
where we have emphasised that $\kappa$ itself depends on derivatives of the potential, as does the left hand side. There is one constraint for each value of $\bar a, \bar b$; these are the promised infinitely many constraints on our theory.

Consistency demands a radius for the circle in the Argand plane greater than zero, a condition that is also in-built in Eq.~\eqref{eq:BonReal} and leads to 
\begin{align}
 \kappa_{\bar a,\bar b}&\leq 1\,.\label{eq:BonKappa}
\end{align}
These are the constraints that this work will use in practice, a more wholistic approach could make use of the more restrictive Eq.~\eqref{eq:BonReal} but Eq.~\eqref{eq:BonKappa} suffices for our exploratory study.

Given the generality of the formulae one can reproduce known results for a single term in the potential, let us take for illustration the bound on the quartic Higgs coupling in the SM. One takes $\bar a=a=2,\bar b=b=0$ hence selecting a particular $\kappa$ and within that $\kappa$ a single term in the sum. This term is the one generated by the quartic $\lambda$ coupling, to be precise, with a SM potential as
\begin{align}
    V_{SM}=-\frac{m_h^2}{4}(v+h)^2+\frac{\lambda}{8}(v+h)^4\,,\label{eq:VSM}
\end{align}
and the aforementioned term in $\kappa_{2,0}$ reads, 
\begin{align}
    \kappa_{2,0}\to\frac{1}{64\pi^2}&\frac{(N-2)!!}{2(N-2)!!}\frac{1}{2}\left|\bar{\mathcal D}_{4,0}\mathcal V_{SM}\right|^2=\frac{1}{(16\pi)^2}\left|3\lambda\right|^2 &  & \Rightarrow & \frac{3\lambda}{16\pi}&\leq 1\,.\label{eq:Ref2pt1}
\end{align}
On the other hand we can borrow the known unitarity bound on $hh$ scattering from~\cite{Lee:1977eg} where the partial wave $a_0$ is given and bounded to be smaller than $1$,
\begin{align}
    a_0(hh\to hh)=&-\frac{3G_Fm_h^2}{8\pi\sqrt{2}}=-\frac{3m_h^2}{16\pi v^2}=-\frac{3\lambda}{16\pi } & |a_0|&\leq 1\,,
\end{align}
which is indeed the very same bound.

The functions $\kappa$ themselves do contain an infinite sum over amplitudes and are the promised observables that have non-local information about the theory. The high energy coefficient of $w^P$ scales for large $P$ as $(\partial ^PV)^2/P!^2$, the connection between the expansion of $V(h)$ and $\kappa (w)$ will be discussed in Section~\ref{sec:Radius}.

For non-renormalisable theories $\kappa$ continues to grow with energy until the constraint is no longer respected and new physics necessarily have to make an appearance. One can estimate this scale $E_*$ as the lowest at which the inequalities saturate, our working definition is
\begin{center}
  \begin{equation}  w_*=\textrm{min}\left(w_{*}^{\bar a,\bar b}\right)\,,\qquad   \kappa_{\bar a,\bar b}(w_*^{\bar a,\bar b})\equiv1\,, \label{eq:CtOffDef} \end{equation}
    \begin{equation}w_*=\frac{s_*}{(4\pi v)^2}=\frac{E_*^2}{(4\pi v)^2}\,.\end{equation}
\end{center}

\subsection{Cross sections and decay rates}\label{sec:XsecDecayR}

We conclude this section by making explicit the connection with physical observables. Indeed for the same in and out state in the unitarity constraint one is squaring the amplitude, integrating over phase space and summing over channels, so that the inelastic cross section for $\bar a$ Higgses and $\bar b$ NGB reads 
\begin{align}
    \sum_{NGB}\sigma_{\bar a,\bar b}^{\textrm{inel}}=\frac{1}{2E_12E_2|v_\textrm{rel}|} \frac{\mathscr P_{2\bar b}\bar a!\bar b!}{((2\bar b-1)!!)^2 \textrm{Vol}_2} \kappa_{\bar a,\bar b}\,,
\end{align}
with Vol$_2=(8\pi)^{-1}$ and $\bar P=\bar a+\bar b=2$, which allows for three possibilities
\begin{align}
    \sigma_{hh}^{\textrm{inel}}=&\frac{8\pi}{s}\kappa_{2,0}(s)\,, &
    \sum_i\sigma_{h\pi_i}^{\textrm{inel}}=&\frac{4 N\pi}{s}\kappa_{1,1}(s)\,, &
    \sum_{ij}\sigma_{\pi_i\pi_j}^{\textrm{inel}}=&\frac{8\pi}{3s}(N^2+2N)\kappa_{0,2}(s)\,.
\end{align} 
The total cross section is obtained by subbing $\kappa\to\kappa+|\mathcal DV/r|^2$ or equivalently not subtracting the $\bar\alpha$ term in Eq.~\eqref{eq:master2}, and e.g.
\begin{align}
    \sigma_{hh}^{\textrm{tot}} =\frac{8\pi}{s}\left(\kappa_{2,0}+\left|\frac{\mathcal D_{4,0} V}{r}\right|^2\right)\,,
\end{align}
and one can write a generalisation for $\bar a+\bar b=2$
\begin{align}
    \sum_{NGB}\sigma_{\bar a,\bar b}^\textrm{tot}&=\frac{4\pi}{s} \frac{\mathscr P_{2\bar b}\bar a!\bar b!}{(2\bar b-1)!!^2}C_{\bar a\bar b}\sum_P^\infty \frac{\alpha_P[V] }{(P-1)!(P-2)!}\left(\frac{s}{(4\pi v)^2}\right)^{P-2}\,,\\
    &=\frac{1}{16\pi s(N-2)!!}\sum_{P'=0}^\infty \frac{\alpha_{P'+2}[V]}{(P'+1)!P'!}\left(\frac{s}{(4\pi v)^2}\right)^{P'}\,.
\end{align}

The above case is for $\bar P=2$ initial particles and gives cross-sections, the $\bar P=1$ case can also be made sense of as a decay rate. In particular with $\bar a=1,\bar b=0$ one obtains the decay rate of the Higgs-like particle to any number of NGB if the threshold of Heaviside functions of Eq.~\eqref{eq:Heavy} is put in place so that the sum is restricted as
\begin{align}
    \Gamma_h&=\frac{1}{2m_h} \frac{\mathscr P_{2\bar b}\bar a!\bar b!}{(2\bar b-1)!!^2V_{\bar a+\bar b}}\kappa_{\bar a,\bar b}(w_h)|_{D^2\to D^2\Theta}\,,\\
    &=\frac{m_h}{256\pi^3{\color{red}}(N-2)!!}\sum_{B}\frac{(2B-1)!!(2B+N-2)!!}{(2B)!(2B-1)!(2B-2)!} \left|\overline{\mathcal D}_{1,2B}\mathcal V\right|^2\left(\frac{m_h^2}{(4\pi v)^2}\right)^{2B-3}\,,\label{eq:Decay}
\end{align}
where $w_h=s_h/(4\pi v)^2=m_h^2/(4\pi v)^2$. Note that for massive particles one can only sum until $2B<m_h/m_\pi$ and hence not a single of these channels happens to be open if one extends this formula to the SM scalar sector. It is nonetheless a good check to reproduce the longitudinal contribution to the decay of a heavy Higgs, one has
\begin{align}
    \Gamma_{h\to W^+W^-}+\Gamma_{h\to ZZ}=\frac{3}{32\pi m_h}\left(\frac{m_h^2}{v}\right)^2+\mathcal O\left(\frac{m_W^2}{m_h^2}\right)
\end{align}
On the other hand straightforward use of Eq.~\eqref{eq:Decay} with $B=1$ for the SM potential  of Eq.~\eqref{eq:VSM} returns
\begin{align}
    \Gamma_h=\frac{m_h}{256\pi^3}\frac{(4\pi v)^2}{m_h^2}\frac{N}{2}|\lambda|^2=\frac{N}{32\pi m_h}\left(\lambda v\right)^2\, \label{eq:HeavyHSMD}
\end{align}
given $m_h^2=\lambda v^2$ this formula returns the SM expression for $N=3$, as it should.

This academic exercise does also serve as a reminder about the equivalence theorem expansion. Results agree up to $m_W^2/m_h^2$ given $m_h$ is the relevant scale of the decay; in general the expansion translates to $m_W^2/s$ which is to say our analysis captures the leading $s$ power of the amplitude. For experimentally accessible energies this term might not be dominant, see ref.~\cite{Bishara:2016kjn} for a study.

For the generalisation to more than two initial-state particles one has that $\kappa$ is proportional to the probability of transition in $\to$ out per unit time, with the proportionality constant a factor of energies and a power of the volume. This type of observables can be relevant in cosmology in e.g. instances in which computing a given particle abundance requires knowledge of transitions with more than two initial particles~\cite{Hochberg:2014dra}. 

\section{Analytic results}
\label{sec:Analytic}

Having derived unitarity constraints, cross sections and decay rates in terms of functions as infinite series on powers of $s$ (or equivalently $w=s/(4\pi v)^2$), this section aims at evaluating them for specific potentials while obtaining analytic results. Analytic is slightly overused in here so let us clarify; for any given potential the central result this work gives explicit $\kappa$ functions and associated cross sections and decay rates analytic in the whole $s$ plane. In some instances the infinite sum these functions are given as can be written in terms of elementary functions; this section will give some of these cases. On the other hand finding $E_*$ as in Eq.~\eqref{eq:CtOffDef} requires finding the inverse of $\kappa$, and this we cannot give in closed form in general. For a given set of parameters this inversion can be solved for numerically; this is what we will refer to as numerical methods in the following, whereas for certain cases and limits the inverse of this function can be given analytically as this section will show.

Each unitarity bound is sensitive to arbitrarily high order covariant derivatives of the potential and holds the promise of identifying non-local properties, yet in a bottom up EFT approach one has that each of these derivatives is an independent variable. In our first attempt at studying the problem we would like to keep the number of parameters finite and manageable while having contributions to arbitrarily high derivatives of the potential to make full use of the constraints. This implies analysing potentials or small-curvature metrics with an infinite covariant Taylor series yet just a few parameters; such a space of functions still gives plenty of choices so for concreteness our case studies will be motivated by physical models. We will attempt to identify results in these concrete examples that can be generalised. 

In this exploration a number of previous works have pointed out at useful theory tools to guide us; in addition to geometry, the study of analytic functions has been shown to be a powerful tool in classifying potentials~\cite{Cohen:2021ucp} and deriving high energy asymptotic behaviour as postulated in~\cite{Falkowski:2019tft}. Let us first outline some necessary elements of analysis to later study cases of increasing complexity.

We will consider the two scenarios below
\begin{itemize}
    \item Partial $\kappa$ results for sums over Goldstone Bosons only. This is simpler to compute yet justified physically. The theory as outlined in Eq.~\eqref{eq:lheft} is a theory of massless NGBs and massive Higgs, in which case for the regime $0<s<m_h^2$ it is only over NGB that we are kinematically allowed to sum. In the application of these results to a theory like HEFT where one gauges part of $SO(N+1)$, this is the $g,g'\to 0$ limit while keeping $v$ fixed.
    \item Fully summed results for both Goldstone and Higgs-like intermediate particles. This is harder to compute, a fact that will reduce our case studies. Particular attention will be paid to illustrate how much is gained by having the full sum as is assessed by determining $E_*$ in each case. 
\end{itemize}

On the technical front a number of results about hypergeometric and gamma functions will be needed and are compiled in appendix~\ref{sec:app}.

\subsection{Radius of convergence}\label{sec:Radius}
It is useful to separate functions into those with and without a singularity in the complex scalar plane. Those without will have a series expansion that converges in the whole plane; those with a singularity also have an expansion but this only converges within a circle on the complex plane of radius given by the distance to the nearest singularity.

One can then start with a rough characterisation of $V(h)$ by a radius of convergence, unbounded from above. The covariant derivative however drives us to another variable as the relevant one to characterise singularities.

For zero curvature one has that Goldstone-direction derivatives read
\begin{align}
    \left(v^2\frac{F'd}{Fdh}\right)^{B}= \left(\frac{v^2}{(h+v)}\frac{d}{d(h+v)}\right)^{B}=\left(\frac{2d}{dI}\right)^B\,,
\end{align}
with 
\begin{align} I=(1+h/v)^2\,,
\end{align}
which gives the Taylor series in terms of $I$ rather than $h$ (in the limit of SMEFT one would have $I=H^\dagger H/\langle H^\dagger H\rangle $) and we note that the vacuum corresponds to $I=1$. It is differential geometry (through covariant derivatives) that has lead us to the group theory result (with I being an `invariant') of $I$ being the relevant variable to study Goldstone emission. Indeed we do know that the linear realisation should admit a regular limit in $I$ as $I\to 0$ where the physical degrees of freedom rearrange from NGBs and Higgs into a $(N+1)$-plet of scalars transforming linearly under EW rotations.

The difference in choice between the variables $h$ and $I$ is that $I(h)$ is analytic while $h(I)$ is not and brings about a branch cut at $h=-v$ which might or might not persist in $V(I)$, the physical magnitude. The statement that can be made in full generality is that an analytic function in $I$ will be analytic in $h$ but for the converse one needs to study the potential. 

The generalisation to the curved case is straightforward
\begin{align}
    \frac{v^2F'}{F}\frac{d}{dh}&\equiv\frac{2d}{dJ}\,, & \int dJ&=\int \frac{2F}{v^2F'}dh\,.
\end{align}

The use of this new variable is that NGB derivatives in Eq.~\eqref{eq:RecRelSol} take the form of ordinary derivatives and complex analysis results can be applied directly to $V(J)$. Let us contemplate the potential as a function of the extended-to-the-complex-plane $J$, where we note that as opposed to $I$, for $J$ we take $J=0$ to be the vacuum (i.e. even in the flat case $J=I-1$),
\begin{align}
    V(h(J))&= \sum_n\tilde V_n J^n \,.
\end{align}
One can find the radius of convergence from the series via the Cauchy-Hadamard theorem,
\begin{align}
    \ell^2=\lim_{n\to \infty} \textrm{sup} \frac{1}{|\tilde V_n|^{1/n}}\,,
\end{align}
yet it can also can be found by the simpler expression
\begin{align}
    \ell^2=\lim_{n\to \infty} \left|\frac{\tilde V_n}{\tilde V_{n+1}}\right|\,,
\end{align}
whenever the limit above exists. 

 For a look at the series in $V(J)$ and its effect on $\kappa(w)$ let us characterise a finite radius with a geometric series $V=(1-J/\ell^2)^{-1}$ which returns $\tilde V_n=(\ell^2)^{-n}$; it follows that the scaling of $\kappa$ for high $P$ and in the NGB direction only would be
\begin{align}
 \kappa&\sim \sum_P\frac{1}{(P!)^2}\frac{(P/2)!^2}{(\ell^2/2)^{P}}w^P\sim\sum_P \frac{w^P}{P!(\ell^2)^{P}}\,, 
 \end{align}
 where we have used $B=P/2$ and Stirling's approximation to derive $(P)!!^2\sim (P)!, (P/2)!^2\sim P!/2^P$ in Eq.~\eqref{eq:master}. This is a result which we can use to find the radius of convergence of the function $\kappa(w)$, 
 \begin{align}
 \ell^2_\kappa&=\lim_{P\to \infty} (P+1)\ell^2=\infty\,.
\end{align}

One has therefore that a finite radius of convergence in $V$ leads to an infinite radius in $\kappa$, i.e. a fully analytic function. One cannot expect therefore for the leading amplitude terms in the high energy limit here studied to lead to a $\kappa$ function with a singularity in the $s$ plane\footnote{Given this better convergence in $\kappa$ one can even ask the question, could a zero radius of convergence $V(I)$ lead to a finite radius for $\kappa$? The unavoidable requirement is a non-zero radius of convergence in $h$, this narrows the question to finding a function $f(x)$ with non-zero radius of convergence around $x=0$ such that $g(y)=f(\sqrt{y}-1)$ has zero radius of convergence around $y=1$.}. The condition for new physics to appear nonetheless is not related to having singularities in $s$; it is an upper bound of $1$ on $\kappa$. To discuss this bound the radius of convergence $\ell^2$ is relevant and will give us an order of magnitude estimate for $w_*$ as we will show in the next subsection.

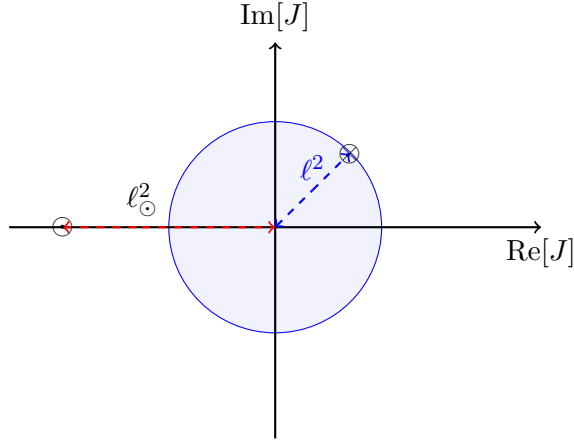
\begin{figure}
    \centering
    \begin{tikzpicture}[scale=0.7]
        \filldraw [blue,fill=blue!5] (0,-3) circle  (2);
        \draw [thick,->] (-5,-3) -- (5,-3) node [anchor=north] {Re$[J]$};
        \draw [thick,->] (0,-7) -- (0,0.5) node [anchor=south] {Im$[J]$};
        \draw [thick,red,dashed,->] (3,-3) -- (4,-3);
        \draw [thick,red,dashed,<-] (0,-3) -- (3,-3)node [anchor=south] {\color{black} $\ell_\odot^2$};
        \draw [thick,blue,dashed,<-] (0,-3) -- (0.7,-2.3) node[anchor=south] {$\ell^2$};
        \draw [thick,blue,dashed,->] (0.7,-2.3) --(1.4,-1.6);
         \draw  (1.4,-1.6) node {$\mathbf{\otimes}$};
        \draw  (4,-3) node {$\odot$};
    \end{tikzpicture}
    \caption{Complex $J$ plane showing the singular point of the potential as $\otimes$ and the would-be symmetry restoring point as $\odot$, each at a distance of $\ell^2$ and $\ell_\odot^2$ respectively of the vacuum.\label{fig:ellandellsun}}
\end{figure}

Lastly, in the current small curvature case one can reduce to the potential the question of linear and non-linear realisations. It takes a simple form illustrated in Fig.~\ref{fig:ellandellsun}, one has that the fixed point (a.k.a. would-be EW symmetry restoring point) $J_\odot$ defined by
\begin{align}
    F(h(J_\odot))\equiv 0\,,
\end{align}
is at a distance $\ell^2_\odot$ in $J$ space whereas the singularity in the potential is at a distance $\ell^2$. In the past the two points have been taken as the same, but in full generality they are distinct and lead to the classification 
\begin{align}
    \ell^2\leq~&\ell_\odot^2\,,& & \textrm{non-linear realisation (HEFT)}\\
    \ell^2>~&\ell_\odot^2\,,& & \textrm{linear realisation (SMEFT)}
\end{align}
and so e.g. the case in Fig.~\ref{fig:ellandellsun} is HEFT-minus-SMEFT.

\subsection{Infinite radius of convergence, a polynomial potential}\label{sec:Poly}

Let us start with a polynomial potential in $I$ in flat scalar space, as one could obtain to finite order in SMEFT. Note that this case has $\ell^2=\infty$ as $V$ is analytic in the whole complex $I$ plane.
In particular we take
\begin{align}
    \mathcal V=\frac{V}{v^4} =
    c\frac{v^{2n-4}}{\Lambda^{2n-4}}I^n\,.
\end{align}

The case of  NGB only summation reads, setting $P=2B$ and using the superindex $\pi$ to denote it is a partial summation,
\begin{align}
    \kappa^\pi_{0,\bar b}=&C_{0,\bar b}w^{(\bar b-4)}\sum_B\frac{(2B+\bar b-1)!!(2B+\bar b+N-2)!!}{(2B)!(2B-1)!(2B-2)!}\left|\left(\frac{2d}{dI}\right)^{B+\bar b/2}c\frac{v^{2n-4}}{\Lambda^{2n-4}}I^n\right|^2 w^{2B}\,,\\
    =&C_{0,\bar b}\left|c\frac{v^{2n-4}}{\Lambda^{2n-4}}\right|^2\sum_{B=1,\neq \bar b/2}^{n-\bar b/2} w^{2B+\bar b-4} \frac{2^{2B+\bar b}(n)!^2(2B+\bar b-1)!!(2B+\bar b+N-2)!!}{(n-(B+\bar b/2))!^2(2B)!(2B-1)!(2B-2)!}\,,\label{eq:GBsPoly}
\end{align}
i.e. a polynomial of highest degree  $w^{2n-4}$, with highest term 
\begin{align}
   \kappa_{0,\bar b}^\pi= &C_{0,\bar b}\left|c\frac{v^{2n-4}}{\Lambda^{2n-4}}\right|^2 w^{2n-4} \frac{2^{2n}(n)!^2(2n-1)!!(2n+N-2)!!}{(2n-\bar b)!(2n-\bar b-1)!(2n-\bar b-2)!}+\mathcal O(w^{p<2n-4})\,.
\end{align}
For illustration let us take the case of $\bar b=2$, and assume a scale $s$ high enough  so that this highest degree dominates, solving for $w_*$ yields
\begin{align}
    \frac{s_*^\pi}{(4\pi)^2}=\left[\frac{(2n-2)!(2n-3)!(2n-4)!(8\pi)^22(2+N)!!}{2^{2n}|c|^2\,(n)!^2(2n-1)!!(2n+N-2)!!3}\right]^{1/(2n-4)} \Lambda^2\,,\label{eq:PolCutGB}
\end{align}
where we note that the large $n$ limit gives a proportionality relation $s\sim n\Lambda^2$, the
 only case with asymptote as $s\to\infty$ is $n= 2$, and $N\to \infty$ seems to bring the cut-off down to zero which would suggest an $N-$dependent normalisation for $c$.
 
\begin{figure}
    \centering
    \includegraphics[width=0.5\linewidth]{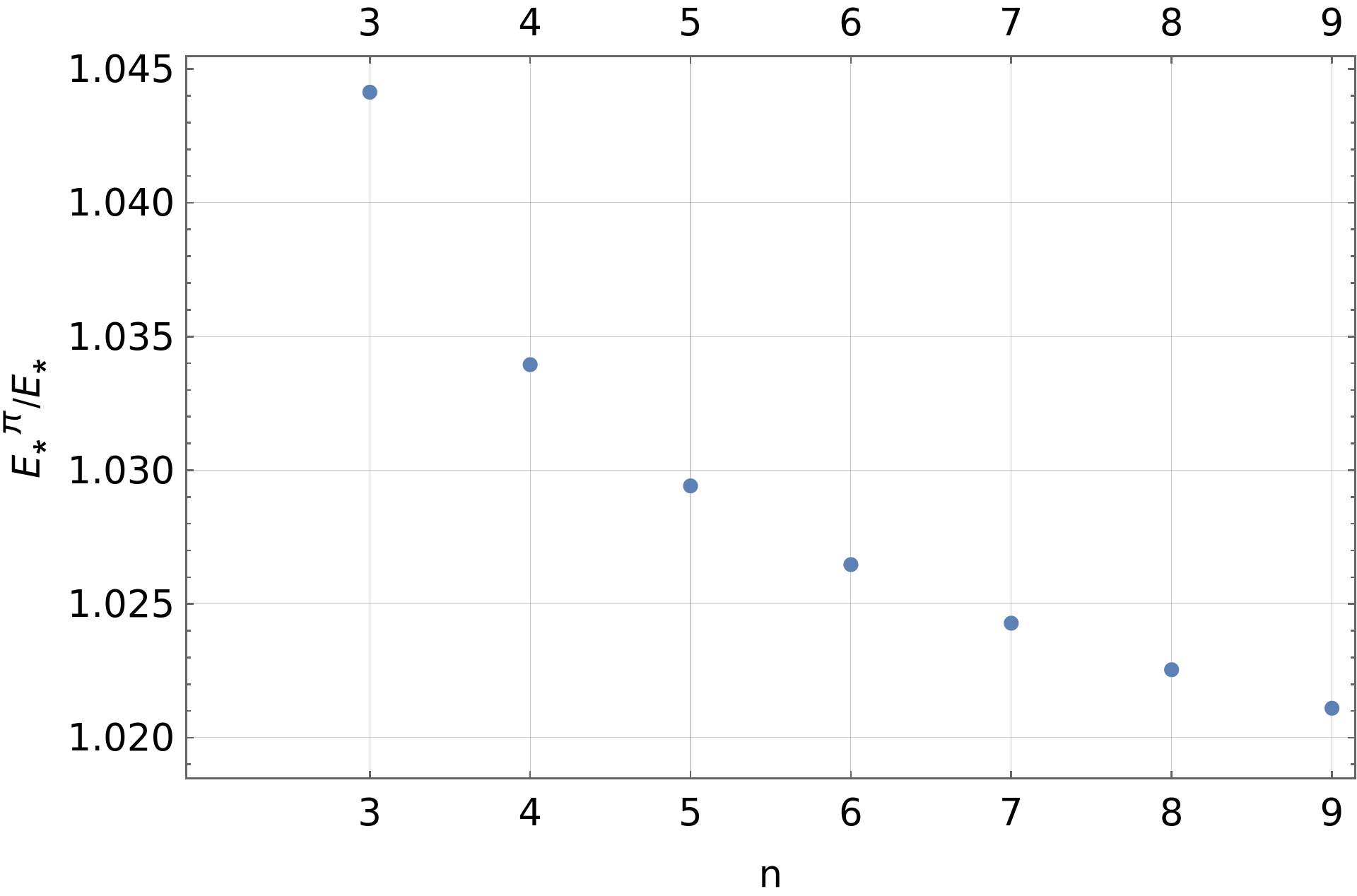}
    \caption{Ratio of cut-off estimates $E_*^\pi/E_*$  vs order of the polynomial $n$ in $I$ (i.e. operator of dimension $2n$) assessing how much the sum over NGB only overestimates the cut-off compared to the sum over NGBs and Higgses which seems to tend to $1$.}
    \label{fig:PolyGBvshGB}
\end{figure}

The choice $V=cv^6I^3/\Lambda^2=c(H^\dagger H)^3/\Lambda^2$, $\bar b=2$ returns a cut-off
\begin{align}
    \frac{s_*^\pi}{(4\pi )^2}=\frac{4\pi\Lambda^2}{|c|\sqrt{45(N+4)}}\label{eq:SMEFT3}\,.
\end{align}
In order to obtain the sum over both NGB and Higgses, we begin by first rewriting
\begin{align}
    [\overline{\mathcal D}_{k,2m}\mathcal V](1)&=\left(\left(\frac{d}{dz}\right)^k\left(\frac{2d}{d(z^2)}\right)^m \mathcal V\right)(1) =c\frac{v^{2n-4}}{\Lambda^{2n-4}}\frac{2^m n!(2n-2m)!}{(n-m)!(2n-2m-k)!}\,,\label{eq:DPoly}
\end{align}
with $z=1+h/v$ and the vacuum at $z=1$.
The  double sum can be turned into a single one as
\begin{align}
    \kappa(w)+\left|\frac{\mathcal D_{2\bar a,2\bar b} V}{r}\right|^2&=\bar C_{\bar a,\bar b} \sum_P\frac{\Gamma^2(P+\bar P-2n)w^{P+\bar P-4}}{\Gamma^2(-n)P!(P-1)!(P-2)!}{}_4F_3\left[\begin{matrix}
        \inu{4}\\
        \ind{3}
    \end{matrix}\,\,;1\right]\,,\label{eq:GBhsPoly}
\end{align}
where we recall $\bar P=\bar a+\bar b$, the function ${}_4F_3$ is the generalised hypergeometric function (see appendix~\ref{sec:app}) with arguments
\begin{align}
   \begin{pmatrix}
       \inu{4}\\
       \ind{3}
   \end{pmatrix} =\begin{pmatrix}
        -P/2&-(P-1)/2&(\bar  b+1)/2 &(\bar  b+N)/2\\
        1/2& (\bar  b+1-\Delta)/2 &(\bar  b+1-\Delta)/2 &
    \end{pmatrix}\,,
\end{align}
and the normalisation constant
\begin{align}
 \bar C_{\bar a, \bar b} &=C_{\bar a,\bar b} \frac{(N-2)!!}{\Gamma(N/2)}\pi 2^{2(\Delta +1)}\Gamma\left[\begin{matrix}
     (\bar b+1)/2 &(\bar b+N)/2\\
     \vec q_3
 \end{matrix}\right]\,.
\end{align} 
where the Gamma function with upper and lower arguments is the ratio of Gamma functions as given in eq~\eqref{eq:GammaProd}.
This form is not straightforward to evaluate since there is a $\infty/\infty$ indeterminate result so let us rewrite the relevant ratio
\begin{align}
    \frac{\Gamma(P+\bar P-2n)}{\Gamma(-n)}=\frac{\Gamma((P+\bar P)/2-n)\Gamma((P+\bar P+1)/2-n)}{\Gamma(-n)\sqrt{\pi}2^{1+2n-P-\bar P}}\,,
\end{align}
as long as $P\leq 2n-\bar P$ we would have a ratio of infinities and one can use Eq.~\eqref{eq:RatNegGam}.
For $P> 2n-\bar P$ the ratio is instead $\Gamma(m)/\Gamma(-n)\sim 1/\infty=0$ and all terms cancel.
The limit value $P=2n-\bar P$ gives the highest $w$ power that appears in $\kappa$; it reads
\begin{align}
\kappa=   {}_4F_3\left[\begin{matrix}
        (\inu{4})_{P=2n-\bar P}\\
        (\ind{3})_{P=2n-\bar P}
    \end{matrix}\,\,;1\right]\frac{n!^2\bar C_{\bar a,\bar b} w^{2n-4}}{4(2n-\bar P)!(2n-\bar P-1)!(2n-\bar P-2)!}+\mathcal O(w^{p<2n-4})\,.\label{eq:GBhsPolyn}
\end{align}
In the same limit as that of Eq.~\eqref{eq:PolCutGB}, the gain in the cut-off bound from summing over all amplitudes rather than NGB only is given by the ratio (setting $\bar a=0$ )
\begin{align}
 \frac{E_*^\pi}{E_*}=   \left({}_3F_2\left[\begin{matrix}
       \inu{3}\\
       \ind{2}
   \end{matrix};1\right]\frac{
     \Gamma((\bar b+1)/2) \Gamma((\bar b+N)/2)
 (N-2)!!\sqrt{\pi}2^{2(n +1)}}{[\Gamma((\bar b+1)/2-n)]^24(2n-1)!!(2n+N-2)!!\Gamma(N/2)}\right)^{\frac{1}{2(2n-4)}}\,,
\end{align}
with \begin{align}
   \begin{pmatrix}
       \inu{3}\\
       \ind{2}
   \end{pmatrix} =\begin{pmatrix}
        -(2n-\bar b)/2&(\bar  b+1)/2 &(\bar  b+N)/2\\
        1/2 &(\bar  b+1-2n)/2 &
    \end{pmatrix}\,.
\end{align}
The plot in Fig.~\ref{fig:PolyGBvshGB} shows this ratio as a function of $n$ for $\bar b=2,\ N=3$; the gain in summing over all channels vs NGB-only is diminishing for increasing power $n$ and for the points shown around the \% level.

\subsection{Finite radius of convergence, a potential with a branch-cut or pole}
\label{sec:finiteR}
To capture the properties of a non-analytic function we use the following potential
\begin{align}
    \mathcal V=\frac{\hat\lambda \ell^{4}}{-\Delta/2}[J/\ell^2+1]^{\Delta/2}\,,\label{eq:potansatz}
\end{align}
which has a singularity or branch cut at a distance of $|\ell^2|$ and a variable overall coupling $\hat \lambda$. The nature of the singularity can be changed with the third free parameter, $\Delta$: for even positive $\Delta$ there is no singularity, for negative even $\Delta$ there is a pole of order $-\Delta/2$, for $\Delta=0$ the potential turns into a logarithm, and for the rest of values there is a branch cut. The three parameters therefore allow for a singularity at a variable distance and a SMEFT limit since the polynomial of the previous section is a sub-case of this potential. The NGB derivative reads
\begin{align}
    \overline{\mathcal D}_{0,2m}\mathcal V&=\frac{-\hat\lambda}{\ell^{2m-4}\Delta/2}\frac{2^m(\Delta/2)!}{(\Delta /2-m)!} =(-1)^{m}\hat \lambda\frac{2^m}{\ell^{2m-4}}\frac{\Gamma(m-\Delta/2)}{\Gamma(1-\Delta/2)}\,,\label{eq:combeha}
\end{align}
where we note that the $\Delta=2\mathbb{Z}^+$ case (i.e. the SM or SMEFT limit) has that for $m\geq\Delta/2+1$ all derivatives cancel and the second form above in terms of $\Gamma$ functions is better suited to study the large $m$ limit.

For both NGB and $h$ derivatives we find only the case with $\ell^2=1$, $J=I-1$ is tractable analytically, with the simple generalisation of Eq.~\eqref{eq:DPoly} as
\begin{align}
    \overline{\mathcal D}_{n,2m}\mathcal V(1)&=\left(\frac{d}{dz}\right)^n\left(\frac{2d}{d(z^2)}\right)^m\mathcal V(1) =
    \frac{-\hat\lambda}{\Delta/2}\frac{2^m(\Delta/2)!(\Delta-2m)!}{(\Delta /2-m)!(\Delta-2m-n)!}\,.
\end{align}

\subsubsection{Sum over Goldstones}\label{sec:AnalyGB}

 In this direction of the double sum we have, for e.g. $\bar a=0$ and $\bar b $ even that the potential of Eq.~\eqref{eq:potansatz} returns 
\begin{align}
    \kappa_{0,\bar b }^\pi+\left|\frac{\mathcal D_{0,2\bar b} V}{r}\right|^2=\underline C_{0\bar b} \frac{|\hat \lambda|^2}{\Gamma^2(1-\Delta/2)}\, \left(\frac{w}{\ell^2}\right)^{\bar b-4}\Gamma\left[\begin{matrix}
        \inu{4}\\
        \ind{5}
    \end{matrix}\right]\left(\,{}_4F_5\left[\begin{matrix}
        \inu{4}\\
        \ind{5}
    \end{matrix};\frac{w^2}{4\ell^4}\right]-1\right)\,,\label{eq:GBHypersum}
\end{align}
with
\begin{align}
    \underline C_{0\bar b}=&C_{0\bar b}\frac{(N-2)!!}{\Gamma(N/2)}\pi 2^{2\bar b+3}\,,&
    \begin{pmatrix}
        \inu{4}\\
        \ind{5}
    \end{pmatrix}&=\begin{pmatrix}
        \frac{\bar b+1}{2}&\frac{\bar b+N}{2}&\frac{\bar b-\Delta}{2}&\frac{\bar b-\Delta}{2}\\
        -\frac{1}{2}&0&0&\frac{1}{2}&\frac{1}{2}
    \end{pmatrix}\,.
\end{align}
This expression gives the cross section 
\begin{align}
   \sum_{ij} \sigma_{\pi_i\pi_j}^\textrm{tot}(s)=\frac{8|\hat \lambda|^2\ell^4v^4}{\Gamma(N/2)\Gamma^2(1-\Delta/2)s^3} \, \Gamma\left[\begin{matrix}
        \inu{4}\\
        \ind{5}
    \end{matrix}\right]\left(\,{}_4F_5\left[\begin{matrix}
        \inu{4}\\
        \ind{5}
    \end{matrix};\frac{s^2}{(2(4\pi v\ell)^2)^2}\right]-1\right)\,.
\end{align}
This cross section is for NGB only and $\pi \,\pi \to n\times\pi$ yet is mediated by a potential $V(h)$ which merits some explanation. The simplest way to realise that a potential on $V(h)$ can lead to NGB scattering in a linear sigma model is to change to Cartesian scalar coordinates instead of $h,\pi$ so that all $N+1$ scalars would feature in the potential; if one insists on staying in $h,\pi$ coordinates, the NGB-only amplitudes coming from a potential are mediated by the kinetic-term coupling of $h-\pi$ and intermediate Higgses which nonetheless have factors of $s/(s-m_h^2)=1+m_h^2/(s-m_h^2)$ and leave behind a contact term even when the Higgs is taken heavy, $s\gg m_h^2$. A special case of this contribution can be found in the longitudinal scattering of W bosons in the SM limit \cite{Lee:1977eg}, where after cancellation of the $\mathcal O(s)$ terms, the next contribution is $G_Fm_h^2\sim \lambda$ with $\lambda$ the quartic coupling.

We can also give the decay rate for the Higgs-like particle using the chain rule  with $\bar b=1$ and adding $\partial J/\partial h$ as
\begin{align}
    \Gamma_{h}=\frac{(4\pi v\ell)^{6}|\hat \lambda v\partial J/\partial h|^2}{256\pi^3 m_h^5(N-2)!!\Gamma^2(1-\Delta/2)}\, \Gamma\left[\begin{matrix}
        \inu{4}\\
        \ind{5}
    \end{matrix}\right]\left(\,{}_4F_5\left[\begin{matrix}
        (\inu{4})_{\bar b=1}\\
        (\ind{5})_{\bar b=1}
    \end{matrix};\frac{m_h^4}{(2(4\pi v\ell)^2)^2}\right]-1\right)\,,
\end{align}
where this formula only applies in the exact global symmetry limit of $m_\pi=0$.

Back to unitarity constraints one has that solving for the cut-off $w_*$ would require inverting a generalised hypergeometric function times a polynomial, something which is not known in terms of elementary functions, even if numerically approachable as shown in the next section. Here instead let us take the high energy limit and property of Eq.~\eqref{eq:asympHFPQ} together with some rewriting to obtain
\begin{align}
    \lim_{w/\ell^2\gg1}\kappa^\pi(w)&= C_{0,\bar b}\frac{\hat \lambda^2}{\Gamma^2(1-\Delta/2)} 2^{2+\Delta} (w/\ell^2 )^{3(\bar b-1)+(N-1)/2-\Delta}\cosh(w/\ell^2 )\,,\\&\simeq C_{0,\bar b}\frac{\hat \lambda^2}{\Gamma^2(1-\Delta/2)} 2^{1+\Delta}(w/\ell^2 )^{3(\bar b-1)+(N-1)/2-\Delta } e^{w/\ell^2 }\,.\label{eq:asympGB}
\end{align}
In this asymptotic limit, finding the cut-off scale amounts to solving $w^p e^w=$const; that is finding the inverse $f(x)$ that satisfies $(f(x))^pe^{f(x)}=x$. This inverse can be given in terms of the $p=1$ solution which is the Lambert or product-log function presented in Eq.~\eqref{eq:Lambert} as:
\begin{align}
    &f(x)= p W_0(x^{1/p}/p)\,.
\end{align}
The estimate for the cut-off scale is 
\begin{align}\label{eq:CutOffAnalytic}
    s_*=E_*^2=&(4\pi v)^2\ell^2\, n_{\bar P}\,W_0\left(\left(\frac{8\pi}{\hat \lambda}\right)^{2/n_{\bar P}}\frac{(\Gamma(1-\Delta/2))^{2/n_{\bar P}}}{(\Omega)^{1/n_{\bar P}}n_{\bar P}}\right)\,,
    \end{align}
    where
    \begin{align}\label{eq:pKdef}
    n_{\bar P}&=3(\bar b-1)+\frac{N-1}{2}-\Delta\,, & \Omega&=\frac{2^{1+\Delta}(\bar b-1)(2\bar b-1)!!}{\bar b!(\bar b-1)!^2(2\bar b+N-2)!!}\,.
\end{align}
The first instances of formulae of the kind \eqref{eq:asympGB} - \eqref{eq:CutOffAnalytic}, which display an exponentiated dependence on energy after summing over infinitely many channels, are to be found in~\cite{Falkowski:2019tft}, followed by \cite{Cohen:2020xca}, where in both cases the sum was performed over Higgs-only intermediate states; some of the focus of this work is a limit (backdoor) which can be spotted in these original formulae. The current formula extends those results by accounting for the initial-state dependence through the dependence on $n_{\bar P}$ and a product-log function in place of the original logarithm.

A few relevant points
\begin{itemize}
    \item The large $\bar b$ or $N$ limit and even the essential singularity limit $\Delta\to-\infty$ yields $s_\star\propto N, \bar b$ or $\Delta$. For a given model $N$ and $\Delta$ are fixed but one might want to study the large value limit of these parameters for a better understanding of the theory. On the other hand $\bar b$ is a variable at our disposal to obtain different bounds and it seems the strongest bounds follow from finite $\bar b$.  
    \item $\ell$ is the scalar distance to the singularity in the potential and the estimate for the cut-off is proportional to it. This offers a straightforward path to decoupling by increasing $\ell$ and suggests this to be the SMEFT limit as indeed it is. Singularities are not therefore exclusive to HEFT, only singularities tied to $v$. 
    \item The limit $n_{\bar P}\to0$ seems singular but in fact returns a logarithm.
    \item The growth of the cut-off with decreasing coupling $\hat\lambda$ is, for positive $n_{\bar P}$ (positive $n_{\bar P}$ can be achieved with large enough $\bar b$), slower than a logarithm but it tends to one asymptotically. This is to say for large enough argument in Eq.~\eqref{eq:CutOffAnalytic} one recovers the logarithm of the original formulae~\cite{Falkowski:2019tft,Cohen:2020xca}.
    \item Even though this is an approximation for the asymptotic behaviour of $\kappa$ and hence implicitly for an original infinite series in $w$, the polynomial case is signalled by $\Gamma(1-\Delta/2)$ which for even positive $\Delta$ sends the cut-off to infinity. This is to say that this formula does not apply any more and one should go back to the polynomial study of Section~\ref{sec:Poly}.
\end{itemize}

\subsubsection{Sum over Goldstones and Higgses}

While we found no extension of the results above for the general potential of Eq.~\eqref{eq:potansatz}, the problem is solvable for the choice of parameters $\ell^2=1$, $J=I-1$ so that
\begin{align}
    \mathcal V=\frac{V}{v^4}=-\frac{\hat \lambda}{\Delta/2} I^{\Delta/2}\,,
\end{align}
provides a closed formula as
\begin{align}
    \kappa(w)+\left|\frac{\mathcal D_{2\bar a,2\bar b} V}{r}\right|^2&=\bar C_{\bar a,\bar b} \sum_P\frac{\Gamma^2(P+\bar P-\Delta)w^{P+\bar P-4}}{\Gamma^2(1-\Delta/2)P!(P-1)!(P-2)!}\,{}_4F_3\left[\begin{matrix}
        \inu{4}\\
        \ind{3}
    \end{matrix}\,\,;1\right]\,,\label{eq:GBhsDil}
\end{align}
where $\bar P=\bar a+\bar b$, the function ${}_4F_3$ is the generalised hypergeometric function with arguments
\begin{align}
   \begin{pmatrix}
       \inu{4}\\
       \ind{3}
   \end{pmatrix} =\begin{pmatrix}
        -P/2&-(P-1)/2&(\bar  b+1)/2 &(\bar  b+N)/2\\
        1/2& (\bar  b+1-\Delta)/2 &(\bar  b+1-\Delta)/2 &
    \end{pmatrix}\,,
\end{align}
and the normalisation constant
\begin{align}
 \bar C_{\bar a, \bar b} &=C_{\bar a,\bar b} \frac{(N-2)!!}{\Gamma(N/2)}\pi2^{2(\Delta +1)}\Gamma\left[\begin{matrix}
     (\bar b+1)/2 &(\bar b+N)/2\\
     \ind{3}
 \end{matrix}\right]\,,
\end{align} 
which is exact to the order in the expansion here considered.

The cross section that follows is
\begin{align}
    \sum_{NGB}\sigma_{\bar a\bar b}^{\textrm{tot}}=\frac{1}{16\pi s(N-2)!!}\frac{\bar C_{\bar a,\bar b}}{C_{\bar a,\bar b}} \sum_P\frac{\Gamma^2(P+\bar P-\Delta)w^{P+\bar P-4}}{\Gamma^2(1-\Delta/2)P!(P-1)!(P-2)!}\,{}_4F_3\left[\begin{matrix}
        \inu{4}\\
        \ind{3}
    \end{matrix}\,\,;1\right]\,.
\end{align}

One can perform the sum on $P$ by further selecting the case $\bar b=0$ and $\Delta=-N+1$; this allows to drop out arguments common to $\inu{}$ and $\ind{}$ to obtain a ${}_2F_1$ function whose form is known explicitly via Gauss summation theorem of Eq.~\eqref{eq:GaussSum}. The resulting form of $\kappa$ is yet another generalised hyperbolic function of argument $2w$ this time
\begin{align}
   \bar b&=0\,,\quad \Delta=-N+1\,, \end{align}
   \begin{align} \kappa(w)+\left|\frac{\mathcal D_{2\bar a,0} V}{r}\right|^2&=C_{\bar a,0}' \frac{w^{\bar P-4}}{\Gamma^2(1-\Delta/2)}\,\Gamma\left[\begin{matrix}
        \inu{3}\\
        (\ind{3})'
    \end{matrix}\right]\,\left({}_{3} F_3\left[\begin{matrix}
        \inu{3}\\
        (\ind{3})'
    \end{matrix};2w\right]-1\right)\,,\label{eq:FullySummed2}
\end{align}
with
\begin{align}
    C'_{\bar a}&=\bar C_{\bar a,0}\frac{2^{N-2}\Gamma(N/2)}{\sqrt{\pi}}\,,
    &
\left(\begin{matrix}
        \inu{3}\\
        (\ind{3})'
    \end{matrix}\right)
    &=\left(\begin{matrix}
       N-1+\bar a& N-1+\bar a &N/2-1/2\\
       0&-1&N-1
   \end{matrix}\right)\,.
\end{align}
As we will discuss in Section~\ref{sec:Dilaton}, this choice of parameters seems incompatible with other considerations like the CFT lower bound on the dimension of operators yet the use we have for it is an explicit case in which to explore how much do constraints improve by summing over all channels. 
With this cautionary reminder let us give here too the cross section that follows
\begin{align}
    \sigma_{hh}^{\textrm{tot}}=\frac{8\pi}{s}C_{\bar a,0}' \frac{w^{\bar P-4}}{\Gamma^2(1-\Delta/2)}\,\Gamma\left[\begin{matrix}
        \inu{3}\\
        (\ind{3})'
    \end{matrix}\right]\,\left({}_{3} F_3\left[\begin{matrix}
        \inu{3}\\
        (\ind{3})'
    \end{matrix};2w\right]-1\right)\,.\label{eq:FullySummed}
\end{align}

Lastly, for this selected potential we can look at the asymptotic result for the cut-off and compare with the NGB only sum, one has that the $w$ dependence tends to 
\begin{align}
\lim_{w\gg1}\kappa(w)\propto   w^{\bar a-4} (2w)^{p_3-q_3'}e^{2w}\sim (2w)^{3(\bar a-1)+3(N-1)/2} e^{2w}\,,\label{eq:asymphGB}
\end{align}
whereas the asymptotic behaviour with these substitutions in the NGB only case of Eq.~\eqref{eq:asympGB}
\begin{align}
    \lim_{w\gg1}\kappa^\pi(w)\propto (w)^{3(0-1)+(N-1)/2+N-1 } e^{w }\sim w^{-3+3(N-1)/2} e^w\,.
\end{align}
Both cases can be fully compared if we select also the (unfortunately non-physical) $\bar a=0$ case where we observe the very same functional form but with double the argument for the full sum. This would imply $\ell^2\to\ell^2/2$ which brings the cut-off down by one-half for $s_*$ or $\sqrt{2}$ for $E_*$. This is a substantial difference and in particular a more relevant one than the polynomial case of Section~\ref{sec:Poly} displayed. Here we have neglected the proportionality constant but these would go inside the Lambert function in Eq.~\eqref{eq:CutOffAnalytic} and hence have a weaker-than-log dependence.

\section{Models that lead to singularities}
\label{sec:Models}

Models with partial or fully complete UV sectors do give rise to potentials and metrics with singularities. Out of the models here studied, some will conform to the potential used for analytic studies of Eq.~\eqref{eq:potansatz} with a choice of $\Delta,\ell^2,\hat\lambda$, some will not; yet for all we find that certain general features derived in Section~\ref{sec:Analytic} are present. The current section with UV realizable scenarios with a few parameters will also enable numerical study and derivation of magnitudes such as the cut-off $E_*$ and hence a test the level of the analytic approximation in Section~\ref{sec:Analytic}. Indeed for theories with scales tied to the EWSB one cannot expect the asymptotic large $s$ approximation to be too accurate given the expectation $s_*/v^2\leq (4\pi)^2$ and this section provides a quantitative assessment of this error.

The main aim of this section is not a quantitative one however, but rather the study of representative UV models and their connection to the parameters characterising the theory at low energies, i.e. the distance to the singularity and the overall coupling. For this purpose we will first study a model with an unbounded distance to the singularity in Section~\ref{sec:Dec}, followed by a second case where this distance is tied-up to the EWSB scale in Section~\ref{sec:NonDec}. These do respectively have a decoupling limit in Section~\ref{sec:Dec} and do not in Section~\ref{sec:NonDec} and further the non-decoupling limit takes us from HEFT to SMEFT. The last category in Section~\ref{sec:Dilaton} will be a type of models that do present a naive decoupling limit from the bottom-up EFT (what we refer to as IR decoupling), different from that in Section~\ref{sec:Dec}, but from the UV perspective or top-down this same limit has a finite cut-off.

\subsection{Decoupling model: an EW singlet}\label{sec:Dec}

Let us take a Higgs doublet $H$ and a scalar singlet $S$ that has taken a VEV, with a $Z_2$ symmetry and coupling to the doublet via the potential 
\begin{align}
    V(H,S)=-m_1^2(H^\dagger H)-\frac{m_2^2}{2}S^2-\frac{\tilde \lambda}{4}S^2(H^\dagger H) +\frac{\lambda}{8}S^4+\frac{\lambda_H}{2}(H^\dagger H)^2\,.
\end{align}
Such an extension is widely used in BSM but in the present geometrical context was first discussed to the best of our knowledge in~\cite{Cohen:2020xca}, here we take a convention for parameters as in~\cite{Alonso:2023upf}.
The scalar takes a VEV provided $m_2^2/\lambda+\tilde \lambda v^2/4\lambda>0$, and has a mass
\begin{align}
    m_S^2&=2m_2^2+\tilde \lambda v^2/2\,.
\end{align}
At low energy it is convenient to define 
\begin{align}
    f_\lambda=&\frac{\tilde \lambda}{2\lambda}\,,& \ell_S^2&=\frac{2m_2^2+\tilde \lambda v^2/2}{\tilde \lambda v^2/2}=\frac{m_S^2}{\tilde \lambda v^2/2}\,,
\end{align}
with $\ell_S$ an educated guess for the singularity distance based on the expectation that it grows with the physical mass of the new particle.
Integrating out $S$ one obtains a non-analytic metric of the form
\begin{align}\label{eq:Smetric}
     \frac{1}{2}\left[\frac{(1+f_\lambda)(I(\underline h)-1)+\ell_S^2+f_\lambda}{ I(\underline h)-1+\ell_S^2}(\partial \underline h)^2+I(\underline h)\,(\partial \pi)^j\hat g_{ij}(\pi)(\partial\pi^i)\right]\,,
\end{align}
and analytic potential 
\begin{align}
    V=&\frac{\ell_S^2+f_\lambda}{8\ell_S^2}m_h^2v^2\left(I(\underline h)-1\right)^2\,,
\end{align}
with $I(\underline h)=(1+\underline h/v)^2$.

This is a case with curvature yet our analysis does still apply if the curvature is small compared to the potential contribution, something we can ensure with a choice of parameters. For this purpose let us give the two invariant cross-sectional curvatures which are
\begin{align}
    v^2R_\pi&=\frac{f_\lambda}{f_\lambda+\ell_S^2}\,,&
    v^2R_h&=\frac{f_\lambda(\ell_S^2-1)}{(f_\lambda+\ell_S^2)^2}\,.    
\end{align}
    Although apparent singularities are easy to spot in the metric, one shall first compute observables, i.e. field-invariant magnitudes, to correctly identify physical singularities. This was done in~\cite{Cohen:2020xca} by looking at the curvature, while here we focus on the potential derived amplitudes.
    
    The metric is not of the form of Eq.~\eqref{eq:lheft} but Eq.~\eqref{eq:underlinehl}, so we bring it to canonical form by the transformation $\underline h(h)$ in Eq.~\eqref{eq:hbarhchange}. In practice we need not solve for this but simply use the chain rule for derivatives as follows
\begin{align}
    \frac{dF}{dh}\frac{1}{F}\frac{d}{dh}=&\left(\frac{d\underline h}{d h}\right)^2\frac{dF}{d\underline h}\frac{1}{F}\frac{d}{d\underline h}=\frac{ I-1+\ell_S^2}{(1+f_\lambda)(I-1)+\ell_S^2+f_\lambda} \frac{d\sqrt{I}}{d\underline h}\frac{1}{\sqrt{I}}\frac{dI}{d\underline h}\frac{d}{dI}\,,\nonumber
    \\=&\frac{2}{v^2}\,\frac{ I-1+\ell_S^2}{(1+f_\lambda)(I-1)+\ell_S^2+f_\lambda}\frac{d}{dI}\equiv\frac{2}{v^2}\frac{d}{dJ}\,.\label{eq:SGBder}
\end{align}

Application of derivatives of the form above on the potential will provide NGB amplitudes and their unitarity bounds, while $V(J)$ will present a singularity. Let us first tackle this problem with the analytic methods of Section~\ref{sec:Analytic} to later use a numerical approach.

While our methods will provide more information, the question we are tackling here is whether or not in the decoupling limit the cut-off does grow linearly with the distance to the singularity, and we will highlight the relevant pieces among the algebra. In this model we have UV input so we do know that the physical cut-off should be around $m_S$, and we can put this against the cut-off derived from unitarity.

In a first attempt let us force-fit the model above into the form of our analytic result in Section~\ref{sec:Analytic}, Eq.~\eqref{eq:potansatz} by approximating $I-1+\ell_S^2$ in the numerator of Eq.~\eqref{eq:SGBder} to $\ell_S^2$. This gives $I\sim (J+\ell_{ap}^2)^{1/2}$ and one can extract 
\begin{align}\label{eq:Sapp}
    \ell^2_{ap}=&\frac{(\ell_S^2+f_\lambda)^2}{2(1+ f_\lambda)\ell_S^2}\,, & \hat\lambda_{ap}&=\frac{m_h^2\ell_S^2}{2v^2(\ell_S^2+f_\lambda)}\,, & \Delta&=1 \,.
\end{align}
Although not a good quantitative approximation this does show the decoupling behaviour in Eq.~\eqref{eq:CutOffAnalytic}; $\ell_S^2$ does send the singularity far away while the coupling tends to a constant and so the analytic approximation points at $s_*\sim (4\pi v\ell_S)^2$.

This quantitative deduction holds for the exact solution of the differential equation that yields
\begin{align}
    I-1&=-\frac{f_\lambda(\ell_S^2-1)}{(1+f_\lambda)}W_{-1}\left[-\frac{\ell_S^2(1+f_\lambda)}{(\ell_S^2-1)f_\lambda}e^{-\frac{J+\ell_S^2(1+f_\lambda)}{(\ell_S^2-1)f_\lambda}}\right]-\ell_S^2 \,,
\end{align}
which, given $W_{-1}(z)$, like $W_{0}(x)$, has a singularity at $z=-1/e$, returns a singularity at a distance,
\begin{align}
   -\ell_{exact}^2= \ell^2_S(1+f_\lambda)+f_\lambda(\ell^2_S-1)\log\left(\frac{f_\lambda (\ell^2_S-1)}{(1+f_\lambda)\ell^2_S}\right)\,,\label{eq:Sappe}
\end{align}
a distance which at large $\ell_S^2$ tends to $\ell_S^2(1+f_\lambda(1+\log[f_\lambda/(1+f_\lambda)]))$  and once more the cut-off grows linearly with $\ell_S$. 
\begin{figure}
    \centering
    \includegraphics[width=0.49\linewidth]{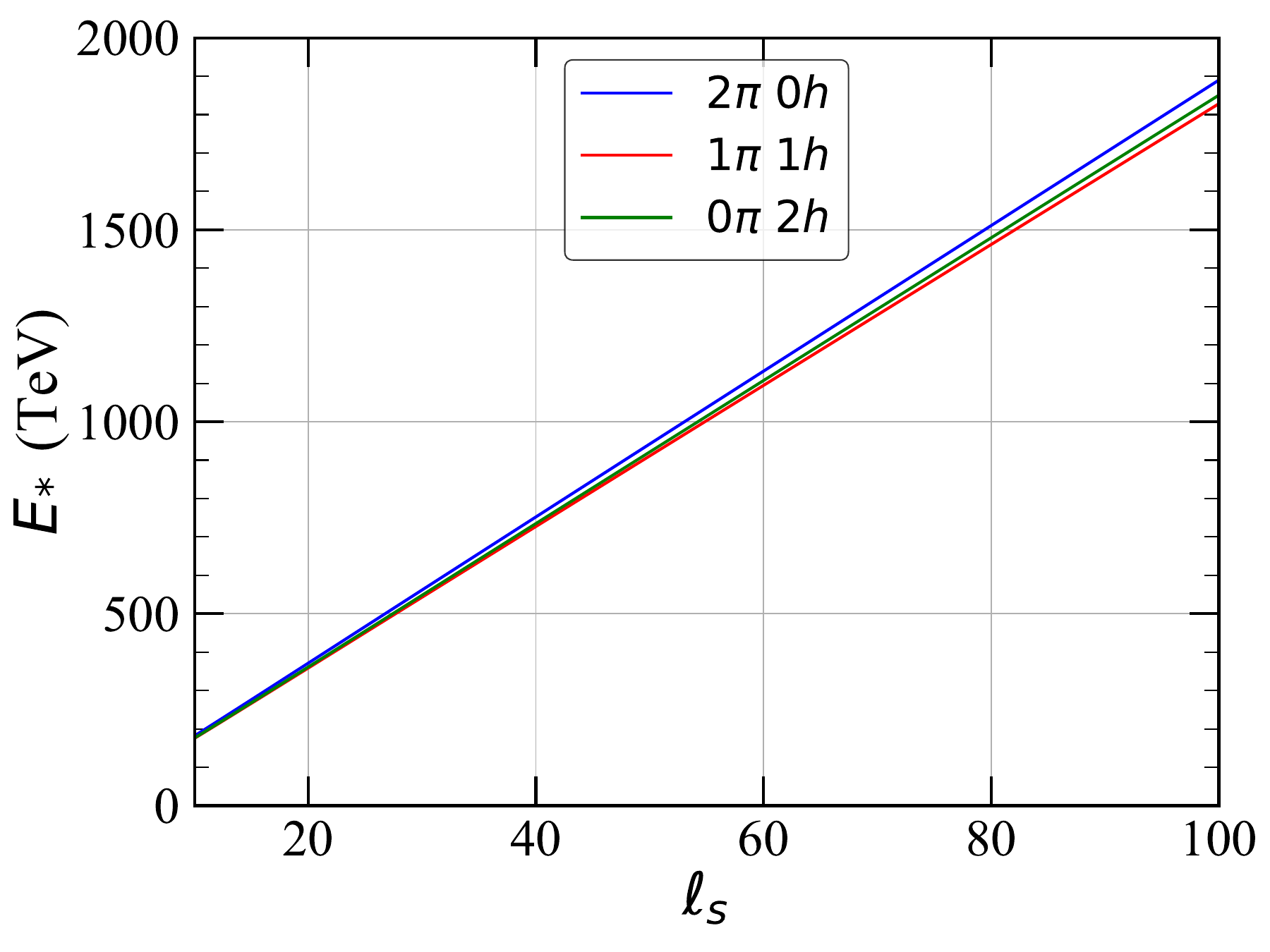}
    \includegraphics[width=0.49\linewidth]{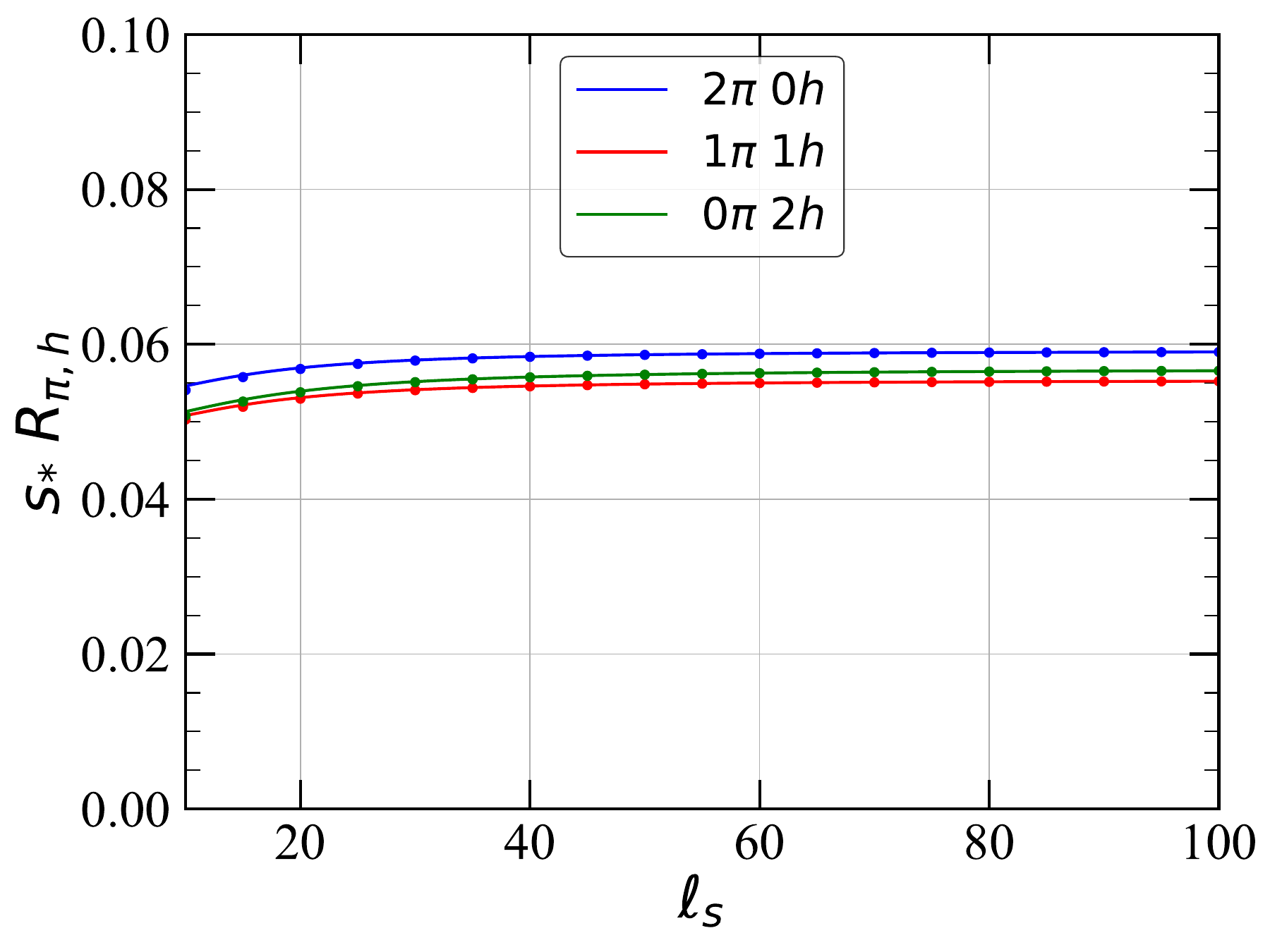}
    \caption{Left panel: unitarity-bound-derived cut-off $E_*$ as a function of $\ell_S$ for $f_\lambda=10^{-5}$ for the decoupling model of Section~\ref{sec:Dec}. Right panel: $s_\star\, R_{\pi,h}$ as a function of $\ell_S$ showing that for our parameter choice curvature effects are subdominant. 
    \label{fig:St}
    }
\end{figure}

\begin{figure}
    \centering
    \includegraphics[width=\linewidth]{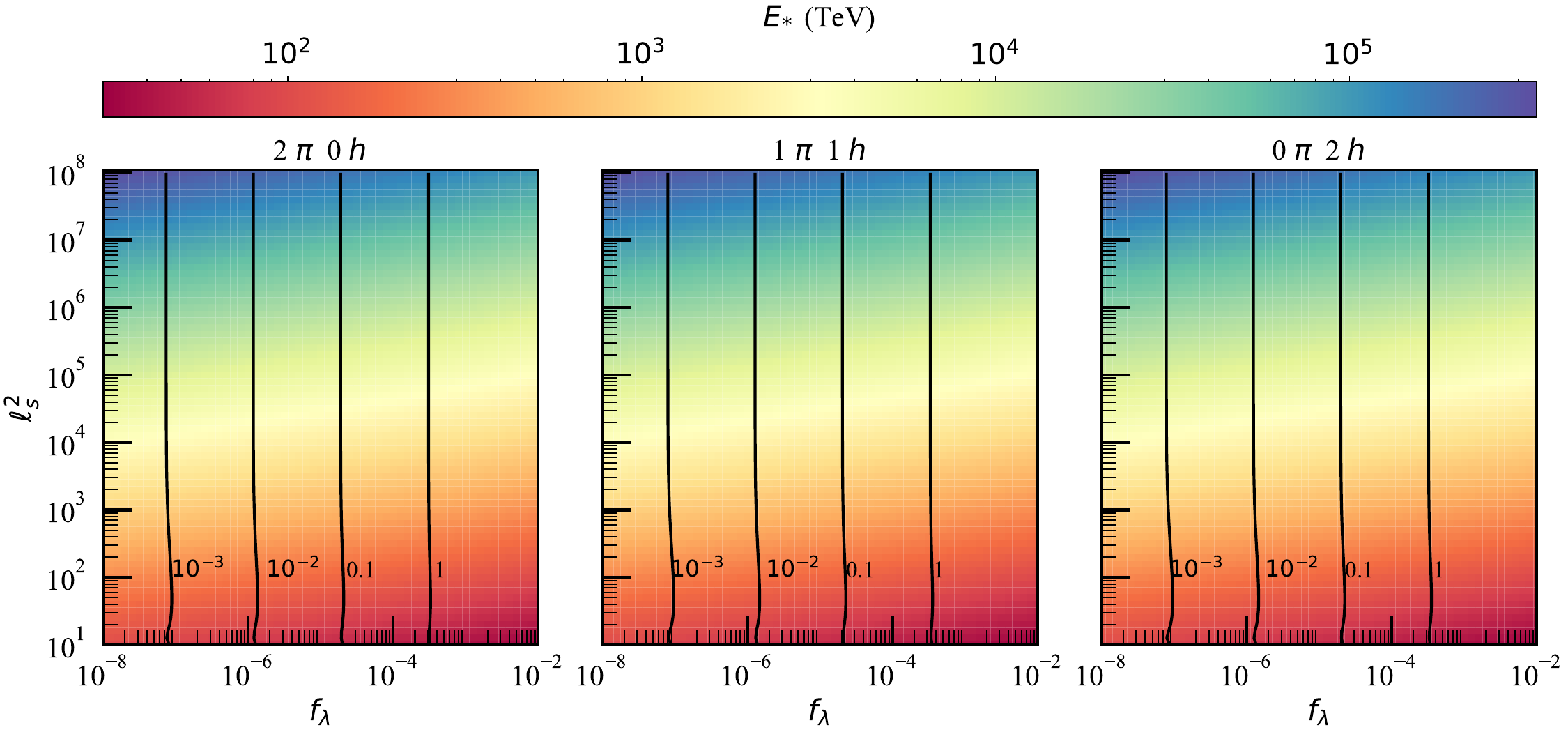}
    \caption{Scalar singlet model unitarity bounds for $2\pi$ (left), $1\pi\,1h$ (middle) and $2h$ (right) incoming particles with $s_\star R_\pi$ labelled in contours. The energy scale (in TeV) for unitarity violation are shown above in the corresponding legends.\label{Splots}}
\end{figure}

 While the analytic approximation returns the predicted behaviour, the present model allows for a complementary approach which we refer to as numeric simply because it is based on the evaluation of the formula in Eq.~\eqref{eq:master2} up to a finite term in the sum and to find its inverse by solving numerically. Figure~\ref{fig:St} shows a plot for the $\bar P=\bar a+\bar b=2$ cases as a function of $\ell_S$, with a value of $f_\lambda=10^{-5}$ chosen so that curvature contributions are kept subleading all the way to the cut-off $E_*$. One can observe that the numerical results do present the linear dependence of $E_*$ with $\ell_S$ anticipated by the approximation, yet quantitatively the approximation is not a good one in the parameter region  plotted (e.g the slope of Fig.~\ref{Splots}, LHS, for $\bar b=2$ is 29.5 while the approximation of Eq.~\eqref{eq:CutOffAnalytic} through Eq.~\eqref{eq:Sapp} returns 5.72 or at best 8.08 if we use Eq.~\eqref{eq:Sappe} ). The main reason why the approximation is not a good one is that for the small value of $f_\lambda$ chosen the metric $hh$ element c.f. Eq.~\eqref{eq:Smetric} tends to one and in Eq.~\eqref{eq:SGBder} the $I$ dependence of the numerator and denominator cancel, whereas in the approximation we neglected the $I$ dependence in the numerator only. 

Finally let us mention that other models with this same decoupling property which could be studied are the log-like potential of~\cite{Kanemura:2021fvp} or UV fermion of~\cite{Cohen:2021ucp}.

\subsection{Non-decoupling model: an EW multiplet}\label{sec:NonDec}

We now turn to the non-decoupling model of Ref.~\cite{Alonso:2021rac}. The model contains the Standard Model doublet together with an additional scalar in the two-index symmetric traceless representation of \(O(N)\). Since this scalar acquires a vacuum expectation value and transforms non-trivially under the electroweak group, expanding around the full-theory vacuum yields the following low-energy field-space metric
\begin{align}
     \frac{1}{2}&\left[(\partial  h)^2+(c_\beta^2I(h)+s_\beta^2)\,(\partial \pi)^j\hat g_{ij}(\pi)(\partial\pi^i)\right]\,,& I=&\left(1+\frac{h}{c_\beta v}\right)^2\,,\label{eq:MetSym}
\end{align}
where for simplicity we have taken the mixing between the low energy Higgs and extra Higgses to zero, keeping it in would complicate the algebra but lead to the same general features.
The potential reads
\begin{align}
     V=\frac{c_\beta^2}{8}m_h^2v^2(I(h)-1)^2\,,\label{eq:PotSym}
\end{align}
while the curvature is
    \begin{align}
        v^2R_\pi&=s_\beta^2\,,&
        v^2R_h&=-s_\beta^2\,.
    \end{align}
    A crucial relation is the connection between EFT parameters in the IR and the mass of the extra heavy Higgs  ($M_{UV}$) in the UV,
    \begin{align}
        \sin2\beta=\sqrt{C_9}\frac{2m_hM_{UV}}{v^2\sqrt{\det(\tilde\lambda)}}\,,\label{eq:UVSym}
    \end{align}
    where $\tilde\lambda$ is a symmetric 2$\times$2 matrix of quartic couplings, $C_9$ is a group dependent factor of order 1; in the minimal $O(4)$ case, it takes the value 8/3 \cite{Alonso:2021rac}.

    The combination of the inputs above implies quantitative application of our methods to this model is limited; one cannot take the zero curvature limit else the mass of the UV particles vanishes, yet large $\beta$ gives large curvature where our results of potential domination over curvature do not apply. Let us consider henceforth a small but non-vanishing $\beta$ with $\beta>1/4\pi$ and hope to describe the model only qualitatively in the following.

    Given that the input on the lower bound on $\beta$ in Eq.~\eqref{eq:UVSym} comes from the UV and hence is external to the EFT approach, let us withhold this information temporarily to explore what the EFT `knows' about the cut-off merely from Eqs.~\eqref{eq:MetSym},\, \eqref{eq:PotSym}.
    
\begin{figure}[t!]
    \centering
    \includegraphics[width=0.65\linewidth]{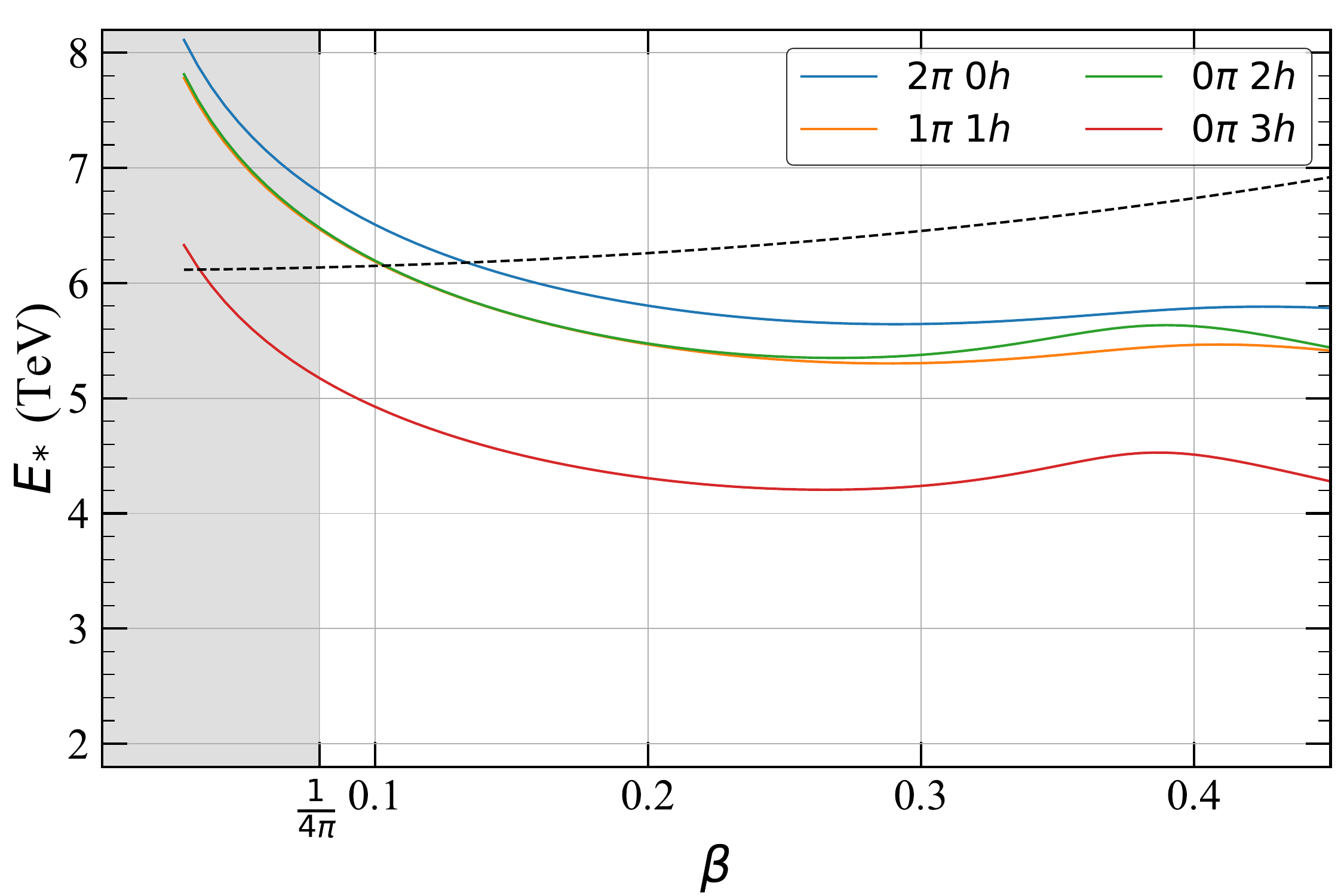}
    \caption{Non-decoupling model unitarity-derived cut-off $E_*$ as a function of $\beta$. Our bounds are a good approximation when $s_\star\, R_{\pi,h} \lesssim 0.1$ which is true for $s_\beta^2 \lesssim 10^{-4}$, the shaded region marks the breakdown of the EFT as given in Eq.~\eqref{eq:UVSym}. The dashed line represents the bounds derived by substituting the approximation Eq.~\eqref{eq:NDapp} in Eq.~\eqref{eq:CutOffAnalytic}. \label{fig:Sym}}
\end{figure}

    The procedure to find the singularity is as before reduced to first solving a differential equation that now reads
  \begin{align}
      \frac{F'}{F}\frac{d}{dh}=\frac{1}{v^2}\frac{2I c^{-2}_\beta}{I+\textrm{tg}_\beta^2}\frac{d}{dI}=\frac{2}{v^2}\frac{d}{dJ}\,.\label{eq:DiffEqSym}
  \end{align}
  One can again force-fit this into the analytically solvable result of Eq.~\eqref{eq:potansatz} by taking the numerator $2Ic_\beta^{-2}\sim 2c_\beta^{-2}$ constant, which yields
  \begin{align}
      \ell_{ap}^2=&\frac{1}{2c_\beta^2}\,, &  \hat\lambda_{ap}&=\frac{m_h^2}{8\ell^2_{ap}v^2}\,, &\Delta&=1\,.\label{eq:NDapp}
  \end{align}
  For small but non-vanishing $\beta$ the order of magnitude estimate provided by this formula is not far off (see Fig.~\ref{fig:Sym}, dashed line) but it fails to capture the limit  $\beta\to 0$ when restricted to the EFT only. The singularity distance stays finite in the $\beta\to 0$ limit for this approximation and so does the coupling, suggesting a finite cut-off when in direct contradiction the limit $\beta\to0$ in Eqs.~\eqref{eq:MetSym},\,\eqref{eq:PotSym} leads to the SM case. One can foresee that his approximation will not give the right result since in this limit it is essential not to neglect the numerator dependence in Eq.~\eqref{eq:DiffEqSym}.
 
  Finding the exact distance to the singularity by solving the exact differential equation on the other hand gives an intermediate relation as
\begin{align}
    J&=\frac{c_\beta^2(I-1)}{2}+\frac{s_\beta^2}{2}\log(I)\,,
\end{align}
where we can observe how the $\beta\to 0$ limit does give an analytic relation between $I$ and $J$ and hence an analytic potential, the SM one. We already know this accounts correctly for this limit but let us carry on to see exactly how, the inversion reads
\begin{align}
    I-1=&\frac{s_\beta^2}{c_\beta^2}W_0\left[\frac{c_\beta^2}{s_\beta^2}e^{\frac{J+c_\beta^2}{s_\beta^2}}\right]-1\,,
    \end{align}
    which gives a radius of convergence
 \begin{align}
    \ell_{exact}^2&=i\pi s_\beta^2 +2s_\beta^2 \log\left(\frac{s_\beta}{c_\beta}\right) - 1 \,.
\end{align}
This distance stays finite for $\beta\to0$ as in the previous, more crude, approximation so where has analyticity (and with it in this case the SM limit) gone? The relation in Eq.~\eqref{eq:CutOffAnalytic} suggests that it is not only the singularity distance, but the coupling $\hat\lambda$ should be studied too; in this case one would have $\hat \lambda \propto s_\beta^2$ so that extrapolating the expression of Eq.~\eqref{eq:CutOffAnalytic} one has $s_*\sim (4\pi v)^2\log(\beta^{-1})$ and the cut-off does increase in this limit pointing at an approach to the SM. 

Turning now to a numerical approach we have a confirmation of the features unveiled in the analytic approach and an order of magnitude quantitative description. Fig.~\ref{fig:Sym} shows a plot for for $E_*$ as a function of $\beta$ for different initial states together with the approximation of Eq.~\eqref{eq:CutOffAnalytic} with the parameters of Eq.~\eqref{eq:NDapp}. In Fig.~\ref{fig:Sym} the UV input of Eq.~\eqref{eq:UVSym} is marked by the shading of the area left of $\beta=(4\pi)^{-1}$.

To summarise, one has that the study of the EFT with no UV information returns a (logarithmically) growing cut-off as $\beta\to0$ and an approach to the SM; this approach however has an unavoidable roadblock around $\beta\sim(4\pi)^{-1}$ in the form of the UV input of Eq.~\eqref{eq:UVSym}. In physical terms, approaching this limit would not yield the SM but will instead produce an ever lighter extra Higgs boson.

This explicit realisation shows the meaning of an EWSB-scale-tied model where the distance to the singularity is bounded from above and there is no SM limit. It does at the same time point to a possibility that we explore next: logarithmic-like decoupling and whether it is ever a UV-complete allowed limit. Exploration of this possibility in this model is obscured by our methods not being quantitatively applicable, so next we turn to a model under full control in our approximation that at the same type features this type of decoupling. 


\subsection{Models with bottom-up decoupling and top-down non-decoupling: the Dilaton EFT}\label{sec:Dilaton}

The dilaton EFT provides an important benchmark within Higgs Effective Field Theory. It describes the dynamics of the pseudo–Nambu–Goldstone boson associated with the spontaneous breaking of scale invariance. This degree of freedom appears whenever an underlying sector exhibits an approximate conformal symmetry that is broken spontaneously, giving rise to a light scalar. An early pedagogical treatment can be found in \cite{Coleman:1985rnk}, and the framework has since been developed extensively.

A near-conformal sector in which conformal symmetry is spontaneously broken can serve as a well-motivated ultraviolet completion of electroweak symmetry breaking, with the familiar electroweak theory emerging as its low-energy limit. In such scenarios, the associated dilaton couples to other massive states with a strength set by their contributions to the trace of the stress–energy tensor, and therefore proportional to their masses. As a result, a light dilaton can naturally reproduce the mass-proportional couplings characteristic of the Standard Model Higgs. This makes it an appealing Higgs impostor candidate \cite{Goldberger:2007zk}, capable of exhibiting SM-like interactions while arising from very different ultraviolet dynamics. Related developments have refined and extended this picture~\cite{Vecchi:2010gj,Chacko:2012sy,Abe:2012eu,Eichten:2012qb,Bellazzini:2013fga,Hernandez-Leon:2017kea,Appelquist:2020bqj,Bruggisser:2022ofg}.

The microscopic origin of the dilaton can be either weakly or strongly coupled. An appealing strongly coupled possibility is provided by non-Abelian gauge theories with many fermion flavours, whose flavour number is taken just below the transition value beyond which an infrared conformal fixed point emerges. In such theories it has been speculated that an approximate realisation of scale invariance develops over a finite range of energies, and that its spontaneous breaking at the confinement scale produces a light dilaton. This picture is supported by two pieces of evidence: first, these theories lie close in flavour number to systems that are genuinely conformal in the infrared; and second, lattice studies of the spectra of two such theories indicate the presence of a light scalar state \cite{LatticeStrongDynamics:2023bqp,LatKMI:2025kti,Fodor:2012ty,Fodor:2015vwa,Fodor:2016pls} which can be interpreted as a dilaton. Furthermore, dilaton EFT has been used to fit lattice data for masses, decay constants and scattering parameters in these gauge theories~\cite{Appelquist:2017wcg,Appelquist:2019lgk,LatticeStrongDynamicsLSD:2021gmp,Appelquist:2022mjb,Golterman:2020tdq,Fodor:2020niv}, while the construction of the EFT and its structural features have also been examined in \cite{Matsuzaki:2013eva,Golterman:2016lsd,Kasai:2016ifi,Golterman:2016cdd,Golterman:2018mfm,Cata:2018wzl,Freeman:2023ket,Zwicky:2023bzk,Zwicky:2023krx,Zwicky:2025moo}. These observations suggest that the dilaton EFT could plausibly arise from a technically natural, UV-complete gauge–fermion theory, which in turn justifies treating it as a meaningful benchmark within HEFT analyses.

We now turn our attention to deriving bounds on the ultraviolet cut-off of dilaton EFT using perturbative unitarity. To do this, we first rewrite the EFT Lagrangian using the geometric HEFT language reviewed earlier in Section~\ref{sec:GeoandPot}. At leading order in derivatives and scale invariance breaking spurions, the Lagrangian for the dilaton EFT can be written as
\begin{align}
    \frac{1}{2}(\partial h)^2+\frac{1}{2}\left(\frac{f_d+h}{v}\right)^2(\partial \pi)^j\hat g_{ij}(\pi)(\partial\pi^i)-V(h)\,,
\end{align}
where the internal, global symmetry is spontaneously broken at the electroweak scale, $v$, and where scale invariance is spontaneously broken at the scale $f_d\ge v$. The dilaton potential is given by
\begin{align}
    V(h)=&\frac{m_h^2(f_d+h)^4}{4(4-\Delta)f_d^2}\left[1-\frac{4}{\Delta}\left(\frac{f_d}{f_d+h}\right)^{4-\Delta} \right]\,.
\end{align}
This potential contains a quartic term with a dimensionless coupling, which is scale invariant and a second term that grows as $h^\Delta$, which breaks scale invariance explicitly. The coefficients of these terms are arranged so that the potential has a global minimum when $h=0$, and has a curvature around this minimum of $m^2_h$, which we identify with the Higgs mass.

Assuming an ultraviolet completion that is a spontaneously broken conformal field theory, perturbed weakly by a scalar operator of scaling dimension $\Delta$, the $h^\Delta$ term captures the leading effect of this deformation within the dilaton EFT. Its appearance, and the corresponding form of the potential, are discussed in Ref.~\cite{Appelquist:2022mjb} and references therein.

The metric develops a conic singularity at the origin unless $f_d = v$, 
a condition we now impose to ensure that the metric curvature vanishes identically ($R = 0$). The Lagrangian that we shall use in the remainder of this section is
\begin{align}
    \mathcal L=&\frac{1}{2}\left[(\partial h)^2+I(h)(\partial \pi)^j\hat g_{ij}(\pi)(\partial\pi^i)\right]-V(h)\,,\\
    \mathcal V=&\frac{V}{v^4}=\frac{m_h^2}{4(4-\Delta)v^2}\left(I^2-\frac{4}{\Delta}I^{\Delta/2}\right)\,.\label{eq:DilLag}
\end{align}

For any scattering channel, unitarity bounds may be computed directly from Eq.~\eqref{eq:master}. When $\bar a+\bar b>2$, however, the analysis simplifies: one may retain only the non-analytic $I^\Delta/2$ term in the dilaton potential of Eq.~\eqref{eq:DilLag}. This is because the analytic $I^2$ term generates no $n$-point contact interactions for $n>4$; moreover, the requirement of at least two intermediate particles for a nontrivial bound implies that $I^2$ does not contribute to $\kappa_{\bar a,\bar b}$ in channels with $\bar a+\bar b>2$.

In the simpler $\bar a+\bar b>2$ case, the results from Section~\ref{sec:finiteR} can be applied, since the non-analytic part of the potential corresponds to Eq.~\eqref{eq:potansatz} with
\begin{align}
\hat \lambda &=\frac{m_h^2}{2(4-\Delta)v^2}\,, &
    \ell^2&=1 \,,
\end{align}
and $\Delta$ being the only variable. With the substitutions above, Eq.~(\ref{eq:GBhsDil}) provides an exact expression for $\kappa$ in the dilaton case, while Eqs.~\eqref{eq:GBHypersum}, \eqref{eq:CutOffAnalytic} can be used as approximations.

\begin{figure}
    \centering
    \includegraphics[width=0.75\linewidth]{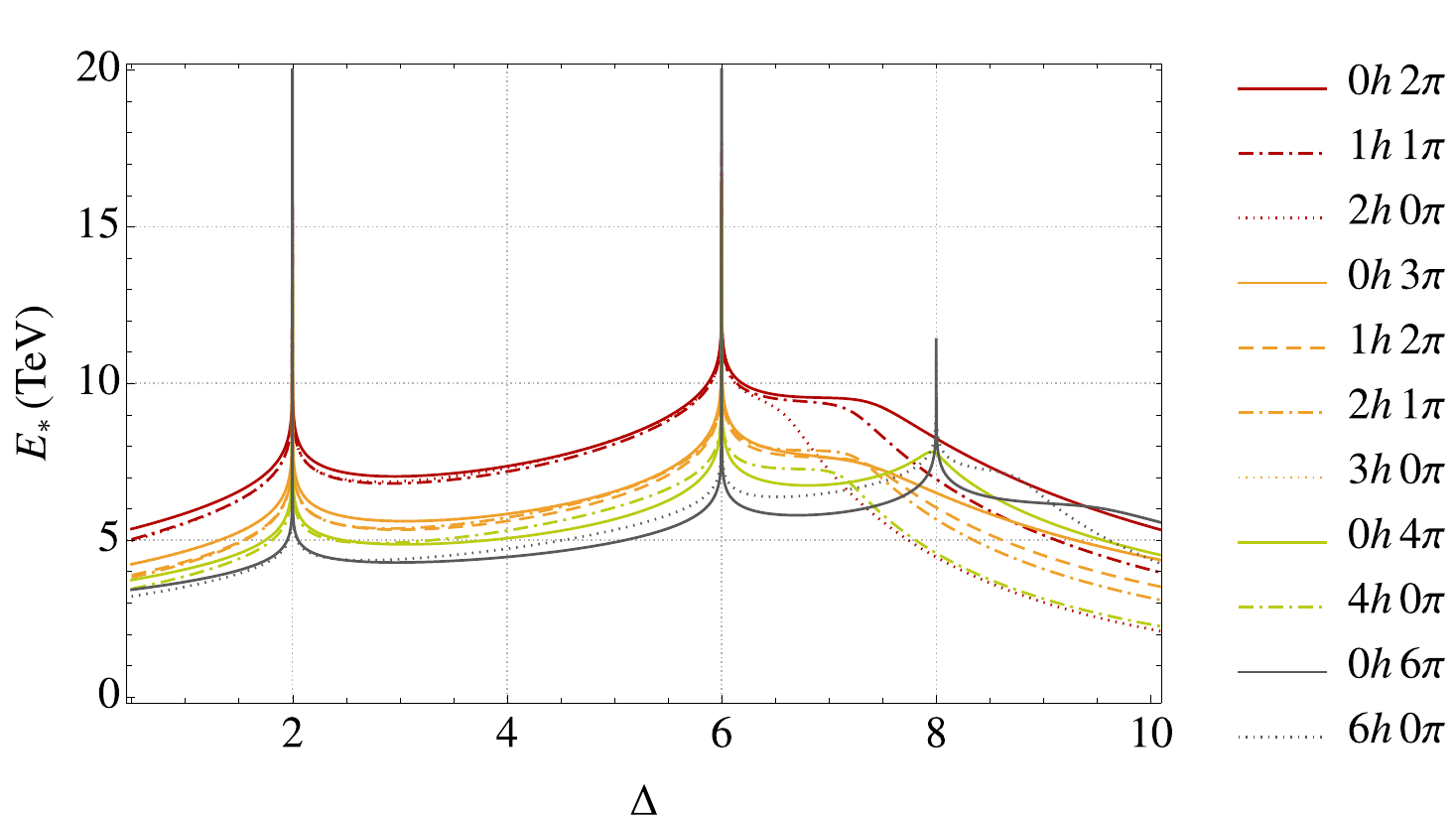}
    \caption{Unitarity bounds on the cut-off of dilaton EFT determined using Eq.~\eqref{eq:master}. The bounds are obtained for different values of the scaling dimension $\Delta$, and for different scattering channels, and are expressed in TeV units.}
    \label{fig:dilatonmain}
\end{figure}

\begin{figure}
    \centering
    \includegraphics[width=\linewidth]{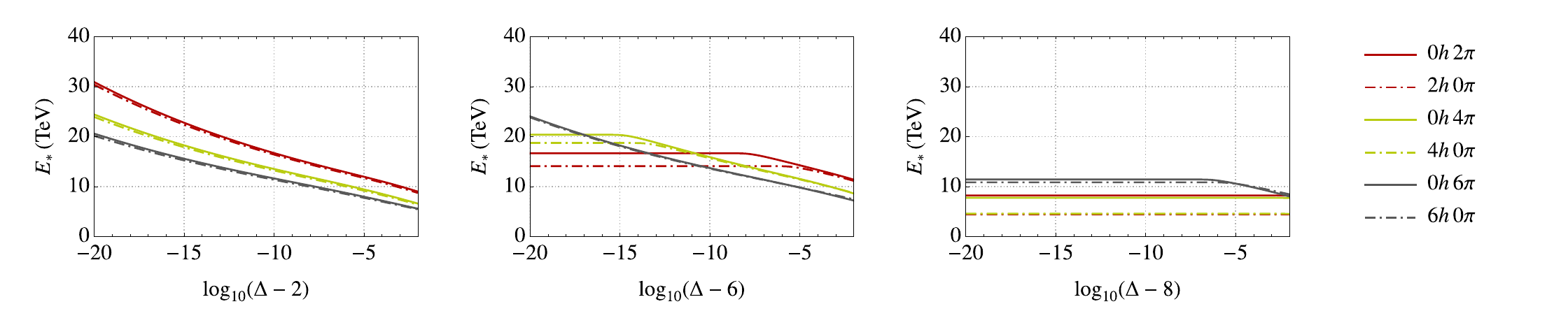}
    \caption{Perturbative unitarity bounds in the regions where $\Delta$ approaches 2, 6 and 8 from above are plotted in the three panels.}
    \label{fig:Dilaton figures}
\end{figure}

In the general case, the construction imposes a fixed distance singularity at $\ell^2=1$, which excludes the would be EW-restoring point of $I=0$ from the scalar manifold. As a consequence the dilaton EFT belongs to the HEFT but not the SMEFT class in general. On the other hand, for particular values of $\Delta$ discussed following Eq.~\eqref{eq:potansatz}, the singularity of the potential is downgraded from a branch cut to a pole, and for even-positive values of $\Delta\neq4$, the potential is analytic.

\begin{figure}[t]
    \centering
    \includegraphics[width=0.9\linewidth]{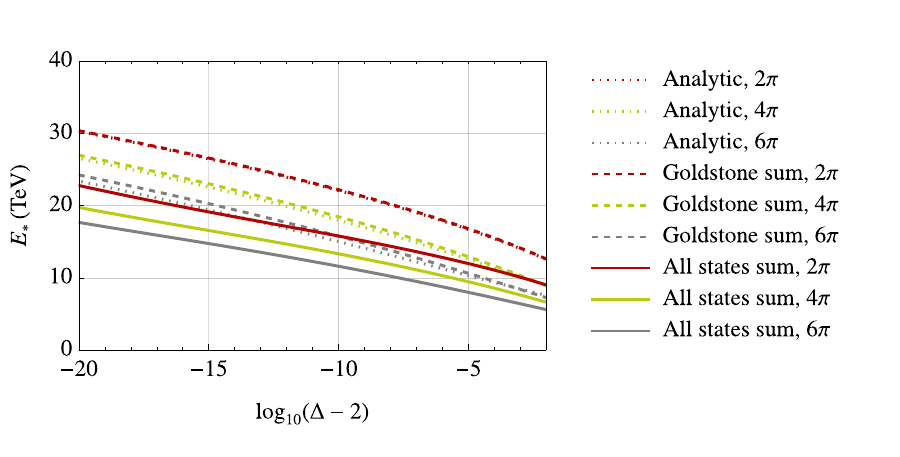}
    \caption{Comparison between unitarity-derived cut-offs in the limit where $\Delta$ approaches 2 from above. Red lines represent the bound obtained from the two Goldstone scattering channel, green lines correspond to the four and grey lines the six Goldstone scattering channel. The dotted lines represent bounds derived using the approximate formula from Eq.~\eqref{eq:aprxw0bound}, dashed lines represent the result from numerically summing over intermediate Goldstone boson states only, and solid lines are the result from summing over all possible intermediate states.}
    \label{fig:bound-compare}
\end{figure}

Figure~\ref{fig:dilatonmain} illustrates the $\Delta$ dependence of the unitarity-derived cut-off $E_*$ from different scattering channels. For $\Delta\neq2$ one has a finite cut-off $E_*$ with a value that depends on the channel as well as $\Delta$. For larger $\Delta$, bounds from channels with larger numbers of particles in the initial state provide the best constraint. Typical values for $E_*$ lie around the $4\pi v\sim 3$~TeV estimate expected of HEFT, but can grow to $6$~TeV around $\Delta\sim7$ for the channels we have shown. 

The most prominent feature is however the divergent points at certain even $\Delta$, where we expect analyticity in the potential as a polynomial in $I$. The case of $\Delta\rightarrow4$ has no divergence or `peak' since the potential tends to a logarithmic Coleman-Weinberg type form in this limit. The peaks of the divergent $\Delta=6,8$ cases are finite as Fig.~\ref{fig:dilatonmain} hints at and Fig.~\ref{fig:Dilaton figures} shows in more detail, unless $\bar a+\bar b>\Delta-2$ in which case they truly diverge. The finite peaks can be made sense of since in that limit one obtains SMEFT but with a fixed scale and hence cut-off, while for $\bar a+\bar b>\Delta-2$ the polynomial potential obtained gives a vanishing contribution to the amplitudes involved which yields an infinite cut-off for the particular channel. The peak at $\Delta=2$ on the other hand returns an infinite cut-off for every single channel and signals the SM limit. 

\begin{figure}[t]
    \centering
    \includegraphics[width=0.6\linewidth]{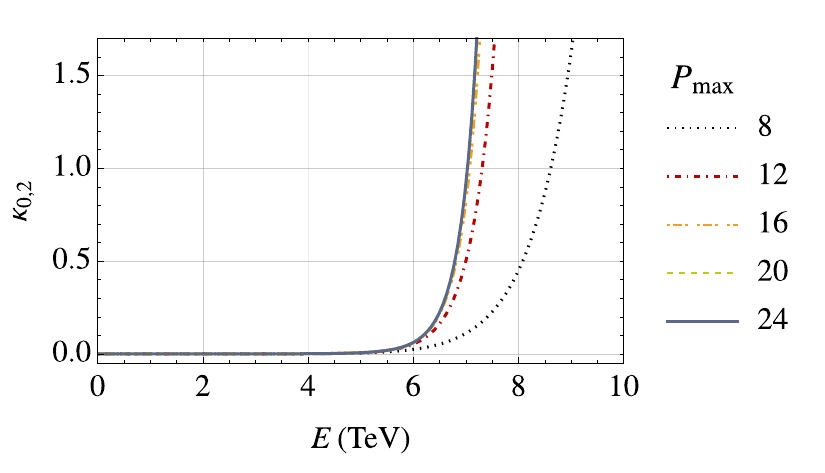}
    \caption{The dependence of the function $\kappa$, as defined in Eq.~\eqref{eq:master}, on energy. Results are shown for the dilaton potential with the representative choice of $\Delta=3$ for the  scattering channel with two Goldstones and zero Higgses in the initial state. The different curves correspond to different values for $P_\text{max}$, the cut-off on the maximum number of particles $P$ retained in the sum over intermediate states.}
    \label{fig:kappavse}
\end{figure}

The peak centred at $\Delta=2$ illustrates our logarithmic decoupling and a possible backdoor to the SM, so we therefore focus on this case. First let us note that, given $\Delta=2+2\epsilon$, and $\epsilon\ll1$ the singular term in the potential reads
\begin{align}
I^{\Delta/2}=I^{1+\epsilon}\simeq I\left(1+\epsilon\log(I)\right)\,,    \label{eq:epslog}
\end{align}
so that there is a branch cut at $I=0$ and the overall coupling is controlled by $\epsilon$.
The cut-off diverges in this limit and we can characterise this divergence using the analytic approximation of Eq.~\eqref{eq:CutOffAnalytic}, which considers only internal states with NGBs but not Higgses and employs an expansion valid for large $s$, and yields
\begin{align}
    s_*=E_*^2=& (4\pi v)^2 n_{\bar P}\,W_0\left(\left(\frac{8\pi}{\hat \lambda}\right)^{2/n_{\bar P}}\frac{(\Gamma(1-\Delta/2))^{2/n_{\bar P}}}{(\Omega)^{1/n_{\bar P}}n_{\bar P}}\right) \,,
    \label{eq:aprxw0bound}
\end{align}
and $n_{\bar P},\,\Omega$ as in Eq.~\eqref{eq:pKdef}. This is an approximation, but one that should be increasingly better as $\Delta$ approaches 2. Fig.~\ref{fig:bound-compare} shows how close this lies to the numerical result. If one takes for concreteness the scattering channel with two Goldstones and zero Higgses so that $n_{\bar P}=2-2\epsilon$, we find 
\begin{align}
    \lim_{\epsilon\to 0} E_*=&\lim_{\epsilon\to 0} 4\pi v \,\sqrt{2} \,\left[W_0\left(\frac{[(8\pi)^2\Gamma^2(\epsilon)(5)!!]^{1/2}}{2[(3)!!(m_h/v)^42^{-1}]^{1/2}}\right)\right]^{1/2}\,,\\ \lim_{\epsilon\to 0}E_*=&4\pi v\,\sqrt{2}\left[\log(1/\epsilon)\right]^{1/2}=4\pi v\,\sqrt{2}\left[\log(2/(\Delta-2))\right]^{1/2}\,,
\end{align}
and we encounter again the logarithmic decoupling limit, but now in an environment where we have analytic control of our approximations. The exact behaviour can be plotted against this prediction and is to be found in Fig.~\ref{fig:bound-compare}.

Regardless of the `speed' of decoupling, what the study above shows is a trajectory that for all finite $\epsilon$ lies in HEFT but at $\epsilon=0$ reaches the SM. This is what we mean by  backdoor to the SM, a way to get to the SM that does not at any point go through SMEFT. This would suggests a configuration as in Fig.~\ref{fig:TheMainPoint}. Furthermore, the decoupling behaviour as $\Delta\rightarrow2$ reflects the approach to a renormalisable limit, and is therefore not expected to be removed by higher-order corrections or additional consistency constraints.

In a similar manner, for values of $\Delta$ approaching an even integer greater than 4 do approach an EFT in the SMEFT category, while these are again signalled in $\Gamma(1-\Delta/2)$, the approximation of Eq.~\eqref{eq:CutOffAnalytic} is not necessarily a good one since the SMEFT limit does have a finite cut-off as given by the polynomial study of Section~\ref{sec:Poly}. This can be appreciated in Fig.~\ref{fig:Dilaton figures}; channels with $\bar a+\bar b\leq\Delta-2$ that present a peak lend themselves to the approximation of Eq.~\eqref{eq:aprxw0bound} but unlike the approximation suggests do have a finite peak. In this sense a feature that starts at $\Delta=8$ and persists for higher $\Delta$ is that certain channels seem not to `notice' the transition to SMEFT with no noticeable feature at even values. 

The issue of thresholds and whether it is accurate to sum over many channels for finite energy is addressed in Fig.~\ref{fig:kappavse}, which shows the convergence of our $\kappa$ functions with the number of internal particles up to which one sums over, for $P_{max}>20$ the lines overlap within the resolution of the plot.

Let us close this section by a general characterisation of backdoors.
For potentials of the polynomial form that belong in SMEFT and serve to characterise analytic functions (i.e. V with radius of convergence $\infty$), the estimate for the cut-off $E_*$ that follows from unitarity is the usual
\begin{align}
    s_*=E_*^2\sim (4\pi \Lambda)^2\,,
\end{align}
with the coefficient of the d-dimensional operator in the potential being $1/\Lambda^{\text{d}-4}$. For non-analytic potentials (i.e. analytic only within a finite radius of convergence $\ell $) with an overall constant controlling the strength of interactions $\hat\lambda$ (and note that through Eq.~\eqref{eq:epslog} the dilaton case has $\lambda\sim\epsilon$), the estimate for the cut-off is instead
\begin{align}
    s_*=E_*^2\sim (4\pi\ell v)^2n_{\bar P} W_0\left(\left[\frac{4\pi}{\hat\lambda}\right]^{2/n_{\bar P}}\right)\,,
\end{align}
with $W_0$ the product-log function which tends to a log for large argument, and $n_{\bar P}$ an order one channel-dependent number. This expression shows how increasing the distance to the singularity $\ell \to \infty$, which at some point brings the EW symmetry-restoring point within the radius of convergence ($\ell>\ell_\odot$ cf. Fig.~\ref{fig:ellandellsun}) and hence transitions to SMEFT, yields a decoupling limit and the SM as the limit theory. Another seemingly valid decoupling limit would be $\hat\lambda\to 0$ with $\ell$ finite, which returns
\begin{align}
    s_*\sim (4\pi\ell v)^2\log\left(\hat\lambda^{-1}\right)\,,
\end{align}
where we note that this can be done with $\ell\leq \ell_\odot$ and so within HEFT. An EFT practitioner who is handed such a potential with $\ell,\hat\lambda$ and takes the limit above would conclude that a decoupling limit wholly within HEFT exists; what we have termed a backdoor. On the other hand all UV complete or semi-complete models here studied reveal such candidate backdoors as a fake; they contain extra resonances of finite and EW-tied masses and are not true decoupling limits. This conflict is the central result of this paper.

\section{Conclusions}
\label{sec:Concl}
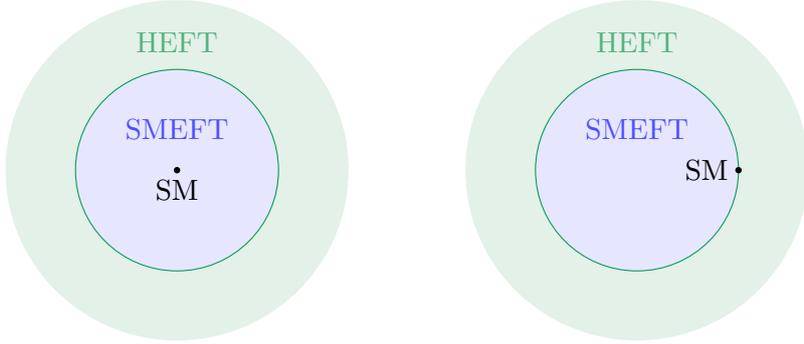
\begin{figure}[t]
    \centering
    \begin{tikzpicture}
        \filldraw[ForestGreen!10,fill=ForestGreen!10] (0,0) circle (64pt);
        \filldraw[ForestGreen,fill=blue!10] (0,0) circle (38pt);
        \filldraw[black] (0,0) node [anchor=north] {SM}circle (1pt) ;
        \draw [blue!70] (0,0.55) node {SMEFT};
        \draw [ForestGreen!70] (0,1.7) node {HEFT};
    \end{tikzpicture}\qquad\qquad
    \begin{tikzpicture}
        \filldraw[ForestGreen!10,fill=ForestGreen!10] (0,0) circle (64pt);
        \filldraw[ForestGreen,fill=blue!10] (0,0) circle (38pt);
        \filldraw[black] (38pt,0) node [anchor=east] {SM}circle (1pt) ;
        \draw [blue!70] (0,0.55) node {SMEFT};
        \draw [ForestGreen!70] (0,1.7) node {HEFT};
    \end{tikzpicture}
    \caption{If there is no other way to obtain the SM than through SMEFT one has the schematic theory space shown in the left panel, whereas if logarithmic decoupling is a viable UV consistent limit one would have a HEFT-backdoor as shown in the right panel.}
    \label{fig:TheMainPoint}
\end{figure}

In this work we have derived the high energy limit of the leading-order amplitudes, arising from the potential, for any number and species of external states in a theory of $N$ Goldstones and one Higgs using geometrical methods.

These amplitudes were then used to derive explicit expressions for inclusive cross sections and decay rates, incorporating an infinite number of decay channels, as well as an infinite set of unitarity constraints, each involving an infinite sum over amplitudes. The results are expressed in terms of geometric derivatives of the scalar potential and are readily applicable to any potential, whether analytic or not. For certain classes of potentials, the expressions can be further summed into hypergeometric functions, and in the high-energy regime a simple approximation yields a unitarity-based cut-off formula that extends the result of Ref.~\cite{Falkowski:2019tft}. While not always quantitatively precise, this approximation captures the qualitative physics and provides useful physical intuition.

These results were put to use in EFT applied to the SM, where the infinite sum over channels is sensitive to non-local properties in field-space and has the potential to distinguish SMEFT (a.k.a. linear realisation) from HEFT (a.k.a. non-linear realisation), where in this context, we use the term HEFT to refer to HEFT-minus-SMEFT, as indicated in Fig.~\ref{fig:TheMainPoint}.

We first studied two ultraviolet benchmark models and their corresponding low-energy effective theories, one in the linear and one in the non-linear realisation of electroweak symmetry, and showed how our framework correctly diagnoses their decoupling and non-decoupling behaviour, respectively. In the decoupling case, the ultraviolet completion contains an additional electroweak singlet scalar beyond the Standard Model field content, whose effects disappear smoothly from low-energy observables. In contrast, the non-decoupling example features an extra scalar transforming nontrivially under the electroweak symmetry group and acquiring a vacuum expectation value that contributes directly to electroweak symmetry breaking alongside the Standard Model Higgs. These well-understood scenarios therefore serve as calibration tests for our approach, before applying it to the case of a dilaton acting as a Higgs impostor.

The dilaton case, which remains under full theoretical control within our approximations across its entire parameter space, may provide a continuous path connecting HEFT directly to the SM, without first passing through SMEFT, provided that the dilaton EFT admits a consistent ultraviolet completion along this path. We have termed this possibility a backdoor, a concept originally identified in \cite{Alonso:2021rac}. In terms of the analytic properties of the theory, the backdoor keeps the distance to the singularity finite but sends the coupling strength to zero, while ordinary or SMEFT-like decoupling sends the distance to the singularity to infinity.

The bottom-up or EFT approach with the unitarity constraints agrees with a backdoor giving a decoupling limit with the cut-off in all channels $E_*\to \infty$, yet our knowledge of the UV model reveals there are resonances at $4\pi v$. This result led us to postulate the question {\it Is there an EFT that can be UV completed with a road to the SM contained in
HEFT?} We do not claim to have a definitive answer to this question. It could be there are some UV considerations that rule out as inconsistent all backdoors akin to swampland arguments ruling out seemingly IR consistent gravity EFTs. If, on the other hand such backdoors do exist, the identification of HEFT with non-decoupling models is not one-to-one: HEFT would also encompass theories that admit a decoupling limit. While the essence of this question is not new, this paper aimed at formulating it as sharply as possible.

\acknowledgments

We would like to thank Aneesh Manohar, Dave Sutherland, Mia West, and Rachel Houtz for valuable comments on the draft, and Tim Cohen for stimulating discussions. J.I. and R.A. are supported
by STFC under grants No. ST/X000745/1 and No. ST/X003167/1. This research was supported in part by the International Centre for Theoretical Sciences (ICTS) for the program -School for Advanced Topics in Particle Physics (SATPP): Selected Topics in Effective Field Theories, code: ICTS/SATPP2024/04.

\appendix
\section{Appendix}\label{sec:app}
Let us compile here the necessary results and special functions. The definition of the generalised hypergeometric function is
\begin{align}
    {}_{k}F_{l}\left(\begin{matrix}
        \vec v_{(k)}\\
        \vec u_{(l)}
    \end{matrix};z\right)\equiv\sum_{n=0}^{\infty}\left(\prod_{i=1}^k\frac{\Gamma(v_{(k)}^i+n)}{\Gamma(v_{(k)}^i)}\right)\left(\prod_{j=1}^l\frac{\Gamma(u_{(l)}^j)}{\Gamma(u_{(l)}^j+n)}\right)\frac{z^n}{n!}\,,\label{eq:HyperDef}
\end{align}
where $v,u$ are vectors of length $k,l$ and components $v_{(k)}^i,u_{(l)}^j$ and the functions converge in the whole plane if $k<l+1$.
The case of a non-positive integer in $\vec v$ means the series terminates since after some $n$ one has $\Gamma(\bar n)/\Gamma(-n)$ with $\bar n\in \mathbb{Z}^+$, $n\in \mathbb{Z}^+,0$ vanishes while for both arguments negative or zero one can use
\begin{align}
    \frac{\Gamma(-\bar n)}{\Gamma(-n)}&=(-1)^{n-\bar n}\frac{n!}{\bar n!}\,,&  n,\bar n&\in \mathbb{Z}^+ \quad \bar n\leq n\,. \label{eq:RatNegGam}
\end{align}
A non-positive integer on $\vec u$ does on the other hand lead to $\Gamma(-n)/\Gamma(\bar n)$ and, after some term, infinite coefficients. This infinity however stems from an overall factor of $\Gamma(u)$ which can be, and in the instances we come across is, removed by simply dividing the whole function by the same factor\footnote{This issue does indeed lead in some textbooks to the definition of functions without the $\Gamma(u)$ factor, the {\it regularised} hypergeometric function.}.

Handy notation is the convention for the gamma function as
\begin{align}
    \Gamma\left[\begin{matrix}
        \vec v\\ \vec u
    \end{matrix}\right]\equiv\frac{\prod \Gamma(v^i)}{\prod \Gamma(u^j)}\,,\label{eq:GammaProd}
\end{align}
whereas subsequent derivations will make frequent use of relations for the $\Gamma$ functions such as Euler's reflection formula, Legendre's duplication formula, Stirling's formula, and the connection with factorials and double factorials.
\begin{figure}[t]
    \centering
    \includegraphics[width=0.48\linewidth]{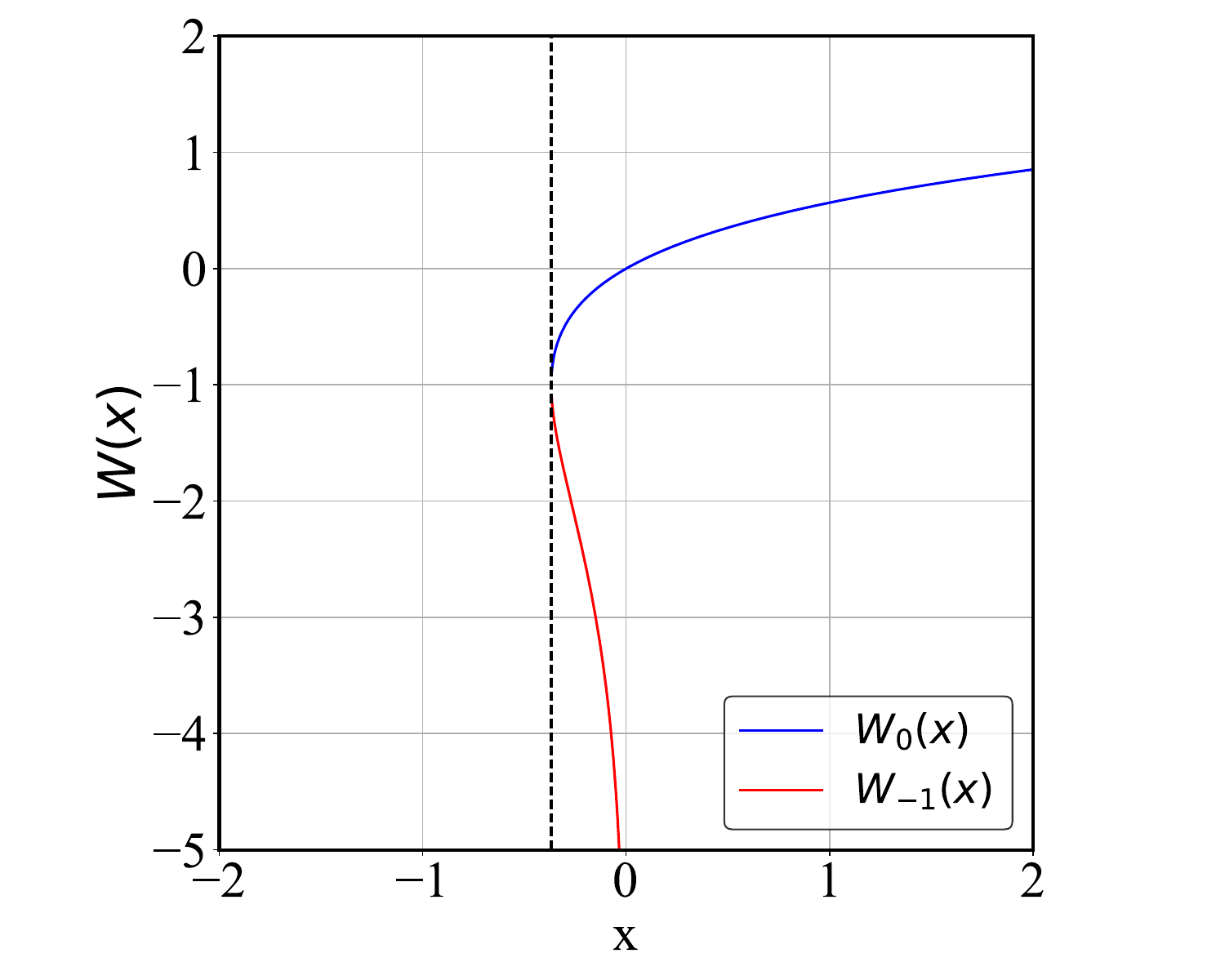}
    \includegraphics[width=0.48\linewidth]{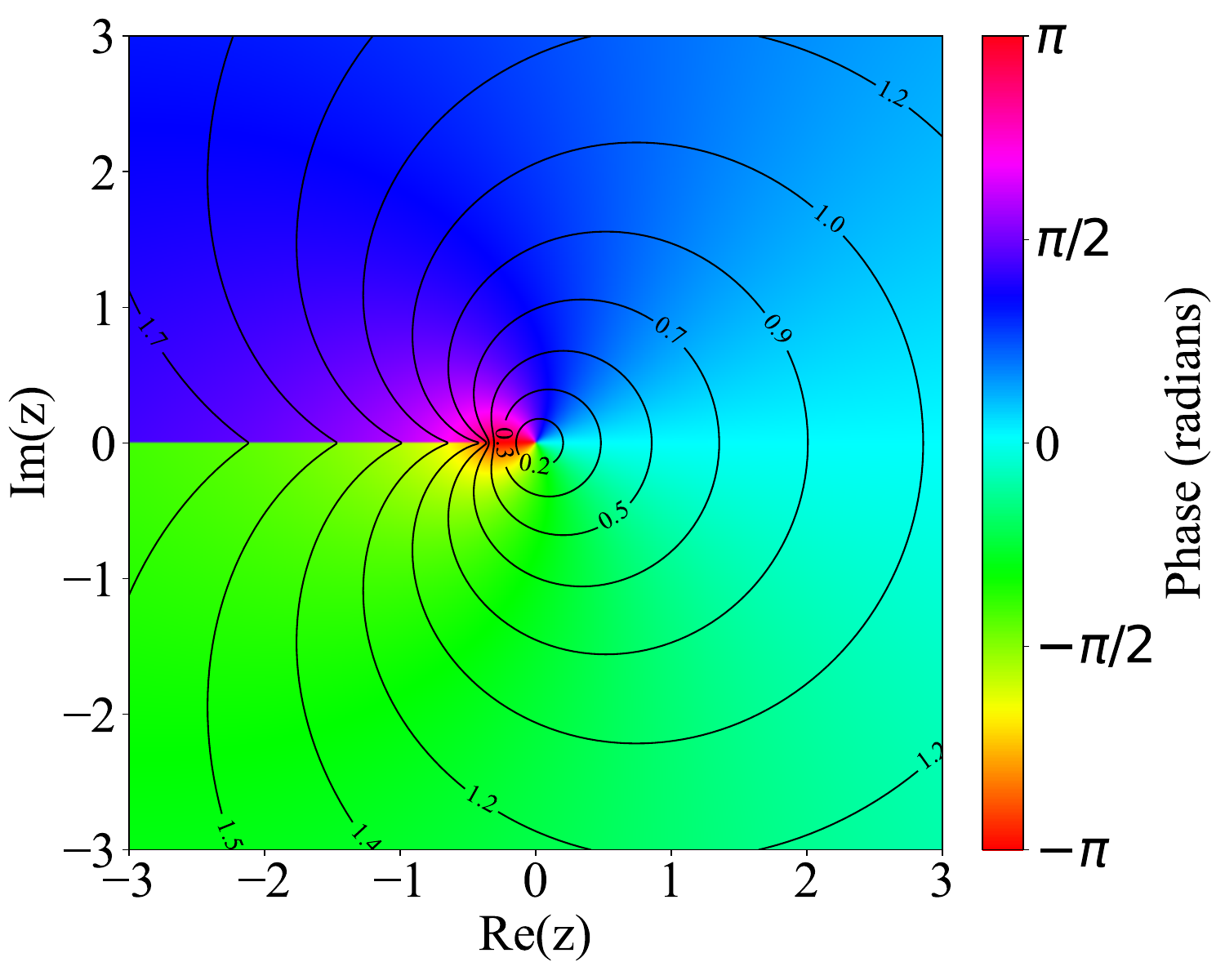}
    \caption{Product-log or Lambert function on the real line in the left hand panel and its extension to the complex plane in the right hand panel with the black solid contour lines represent the contours of $\lvert W_0(z)|$ and the colour map represents the phase of $W_0(z)$.}
    \label{fig:Lambert}
\end{figure}

The asymptotic behaviour of hypergeometric functions is given by the large $n$ coefficients which tend to
\begin{align}
   \left(\prod_{i=1}^k\frac{\Gamma((u_{(k)}^i+n)}{\Gamma(u_{(k)}^i)}\right)\left(\prod_{j=1}^l\frac{\Gamma(u_{(l)}^j)}{\Gamma(u_{(l)}^j+n)}\right)\to     \Gamma\left[\begin{matrix}
        \vec v_{(k)}\\ \vec u_{(l)}
    \end{matrix}\right][(n-1)!]^{k-l} n^{\sum v^i_{(k)}-\sum u^j_{(l)}} \,,
\end{align}
so that in particular the $k=l$, $z\gg1$ limit
\begin{align}
   \lim_{z\gg 1} {}_kF_{k}\left(\begin{matrix}
        \vec v_{(k)}\\
        \vec u_{(k)}
        \end{matrix}\right) = \Gamma\left[\begin{matrix}
        \vec v_{(k)}\\ \vec u_{(k)}
    \end{matrix}\right] z^{\sum (v-u)^i}e^z\,.\label{eq:asympHFPQ}
\end{align}
Another property of hypergeometric functions which we will use is Gauss summation theorem which reads
\begin{align}
    {}_2F_{1}\left(\begin{matrix}
         v& v'\\
        u
    \end{matrix};1\right)=\frac{\Gamma(u)\Gamma(u-v-v')}{\Gamma(u-v)\Gamma(u-v')}\,.\label{eq:GaussSum}
\end{align}

Finally the Lambert or product-log function which will appear at several different places in the analysis is defined as
\begin{align}
    W_0(x)\quad  \textrm{ is such that}\quad W_0(x) e^{W_0(x)}=x\,, \label{eq:Lambert}
\end{align}
and it has a branch cut at $x=-1/e$. The solution to the defining equation of $W_0$ is not unique since the function $y e^y$ is surjective but not injective and another solution provides $W_{-1}$ which is depicted in Fig.~\ref{fig:Lambert}.




\bibliographystyle{JHEP}
\bibliography{biblio.bib}
\end{document}